\documentclass[aps,pre,showpacs,noshowkeys,amsmath,amssymb,amsfonts,superscriptaddress,longbibliography,reprint]{revtex4-1}
\usepackage[english]{babel}

\usepackage{graphicx}
\usepackage{bm}
\usepackage{physics}
\usepackage{mathtools}
\usepackage{gensymb}

\setcitestyle{super}
\usepackage{caption}
\usepackage{subcaption}
\DeclareCaptionLabelSeparator{bar}{~\rule[-0.4ex]{0.2ex}{1em}~}
\DeclareCaptionLabelFormat{subfor}{\textbf{#2}}
\newcommand{\com}[1]{\textcolor{black}{#1}}
\usepackage{xcolor}
\definecolor{UBcolor}{HTML}{007CC1}
\usepackage[colorlinks=true,pdfnewwindow=true,linkcolor=UBcolor,citecolor=UBcolor,urlcolor=UBcolor,breaklinks=true,linktocpage]{hyperref}
\usepackage[all]{hypcap}
\usepackage[nameinlink,capitalise]{cleveref}
\crefname{SI section}{SI Section}{SI Sections}
\Crefname{SI section}{SI Section}{SI Sections}
\begin{document}

\title{Active transport in complex environments}
\thanks{Preprint of a chapter in the book \textit{Out-of-Equilibrium Soft Matter: Active Fluids}, to be published by the Royal Society of Chemistry.}
\author{Alejandro Mart\'{i}nez-Calvo}
\affiliation{Fluid Mechanics Group, Gregorio Mill\'{a}n Institute, Universidad Carlos III de Madrid, 28911 Legan\'{e}s, Spain}
\affiliation{Department of Chemical and Biological Engineering, Princeton University, Princeton, NJ 08544, USA}
\affiliation{Princeton Center for Theoretical Science, Princeton University, Princeton, NJ 08544, USA}

\author{Carolina Trenado-Yuste}
\affiliation{Mathematics Department, Gregorio Mill\'{a}n Institute, Universidad Carlos III de Madrid, 28911 Legan\'{e}s, Spain}
\affiliation{Lewis-Sigler Institute for Integrative Genomics, Princeton University, Princeton, NJ 08544, USA}

\author{Sujit S. Datta}
\email{ssdatta@princeton.edu}
\affiliation{Department of Chemical and Biological Engineering, Princeton University, Princeton, NJ 08544, USA}

\date{\today}

\begin{abstract}
The ability of many living systems to actively self-propel underlies critical biomedical, environmental, and industrial processes. While such active transport is well-studied in uniform settings, environmental complexities such as  geometric constraints, mechanical cues, and external stimuli such as chemical gradients and fluid flow can strongly influence transport. In this chapter, we describe recent progress in the study of active transport in such complex environments, focusing on two prominent biological systems---bacteria and eukaryotic cells---as archetypes of active matter. We review research findings highlighting how environmental factors can fundamentally alter cellular motility, hindering or promoting active transport in unexpected ways, and giving rise to fascinating behaviors such as directed migration and large-scale clustering. In parallel, we describe specific open questions and promising avenues for future research. Furthermore, given the diverse forms of active matter---ranging from enzymes and driven biopolymer assemblies, to microorganisms and synthetic microswimmers, to larger animals and even robots---we also describe connections to other active systems as well as more general theoretical/computational models of transport processes in complex environments. 

\end{abstract}

\maketitle

\section{Introduction}\label{sec:intro}
A key feature of many living systems, such as bacteria and eukaryotic cells, is their ability to actively self-propel. Indeed, cellular motility underlies numerous critical biological processes, such as infection and biofilm formation by bacteria, cancer metastasis, tissue repair and wound healing, and morphogenesis of new tissues and organs. As a result, motility behaviors have been extensively investigated for diverse classes of cells, both in isolation and in groups, in uniform environments such as liquid cultures or flat Petri dishes. These studies have provided tremendous insight into the biological and physicochemical mechanisms that engender and regulate cellular motility. 

However, cells often inhabit complex and non-uniform spaces---e.g., gels and tissues in the body, or soils and sediments in the environment---that impose additional geometric constraints, mechanical cues, and other external stimuli such as chemical gradients and fluid flow. These factors can fundamentally alter cellular motility, hindering or promoting active transport in unexpected ways, and giving rise to fascinating behaviors such as directed migration and large-scale clustering. These phenomena are not only biologically important, but have consequences for industrial and environmental processes ranging from chemical sensing/delivery to sustaining plant growth in agriculture. Moreover, studies of these phenomena motivate new ways of thinking about transport processes---in many cases, challenging current understanding based on studies of either active matter in uniform settings or ``passive'', thermally-equilibrated matter in complex environments. Biological systems are structurally messy, rheologically complex, and are inherently multicomponent, disordered, and out of equilibrium, and yet, they robustly self-assemble and function in a coordinated manner. Thus, studies of active transport in complex environments could yield new insights for the control of soft and active materials more broadly. It could also inspire the development of new forms of adaptive and multifunctional matter that are capable of recapitulating and even going beyond the capabilities of living systems.

In this chapter, we describe recent progress in studies of active transport in complex environments. We focus on two prominent biological systems---bacteria and eukaryotic cells---as archetypes of active matter; however, where relevant, we also reference works on other forms of active matter as well as more general theoretical/computational models. Our goal is not to present a comprehensive overview of all the literature in this field, but rather, to highlight some areas of research whose growth has been particularly rapid. Along with describing research findings, within each section, we describe specific open questions and promising avenues for future research. Finally, motivated by the diverse forms of active matter---ranging from enzymes and driven biopolymer assemblies, to microorganisms and synthetic microswimmers, to larger animals and even robots---we conclude this chapter by highlighting general principles connecting active systems across vastly different length and time scales.

\section{Bacteria}\label{sec:sec2}
Bacteria are an archetype of active microswimmers that can move using a variety of different mechanisms. Indeed, the ability of many bacteria to self-propel was noted in their initial discovery in 1676: upon peering into a drop of pond water with a microscope, Antony van Leeuwenhoek remarked \textit{``that no more pleasant sight has ever yet come before my eye than these many thousands of living creatures, seen all alive in a little drop of water, moving among one another, each several creature having its own proper motion."}~\citep{bergsciam} Over the centuries since, researchers have extensively characterized the motility of diverse species of bacteria, typically focusing on cells in uniform environments such as in liquid cultures or near flat surfaces. 

A common mechanism by which many bacteria self-propel in fluid is using one or more flagella---slender, actively-moving $\sim5~\mu$m-long appendages attached to the cell body surface~\citep{berg}. For the canonical examples of \textit{Escherichia coli} and \textit{Bacillus subtilis}, as well as many other species of bacteria, multiple helical flagella rotate synchronously as a bundle, enabling the cell to ``push'' surrounding fluid~\citep{lauga2009hydrodynamics} and swim along directed, ballistic ``runs'' of mean speed $\bar{v}_{\rm{r}}\sim25~\mu$m/s. Runs do not proceed indefinitely; instead, they are punctuated by reorientations (``tumbles'') caused by one or more of the flagella rotating in the opposite direction, which forces the flagellar bundle to splay out and reorient the cell body, before rapidly coming back together and pushing the cell along another run in a different direction. Runs therefore have a mean length $\bar{\ell}_{\rm{r}}\sim50~\mu$m. Compared to runs, which last on average $\bar{\tau}_{\rm{r}}\sim2$ s, tumbles are effectively instantaneous, lasting on average $\bar{\tau}_{\rm{t}}\sim0.1$ s. Hence, over large length and time scales, this run-and-tumble motion can be modeled as a random walk with an active translational diffusion coefficient given by $D_{\rm{b}}\sim\bar{\ell}_{\rm{r}}^{2}/\bar{\tau}_{\rm{r}}\sim1000~\mu$m$^{2}$/s. While other species of bacteria self-propel using other motility mechanisms, resulting in e.g., run-reverse or run-reverse-flick motion instead, these dynamics can again be modeled as random walks with active translation diffusion coefficients that are determined by the features of single-cell propulsion ~\citep{Xie:2011,Taktikos:2013,theves2013bacterial,Taute:2015}. 

While this description enables bacterial transport to be predicted in uniform liquids, in many real-world settings, bacteria must navigate complex environments such as gels and tissues in the body, micro- and meso-porous foods, and soils, sediments, and subsurface formations in the environment. In some cases, bacterial transport can be harmful to humans: for example, during an infection, pathogens squeeze through pores in tissues and gels and thereby spread through the body~\citep{balzan,chaban,pnas,harman,ribet,siitonen,lux,oneil}---potentially leading to the formation of virulent biofilm communities that are recalcitrant to treatment. Similarly, during meat spoilage, pathogenic bacteria squeeze through pores in tissue and spread in contaminated meat~\citep{Gill1977,Shirai2017}. 

Bacterial spreading can also be beneficial: for example, a promising route toward cancer treatment relies on engineered bacteria penetrating into tumors and delivering anticancer agents ~\citep{thornlow,Toley:2012}. In agriculture, rhizosphere bacteria move through soils to help sustain and protect plant roots, which impacts crop growth and productivity ~\citep{dechesne,souza,turnbull,watt,babalola}. Furthermore, in environmental settings, bioremediation efforts often seek to apply motile bacteria to migrate towards and degrade contaminants trapped in porous groundwater aquifers ~\citep{roseanne18,Adadevoh:2016,ford07,wang08}. Moreover, motility enables bacteria to adopt symbiotic relationships with other organisms in natural environments, helping to promote ecological stability and even shaping evolutionary processes ~\citep{raina2019role}. 

However, despite their potentially harmful or beneficial consequences, there is an incomplete understanding of how bacteria move in such complex environments---which inherently have heterogeneous structures that impose physical obstructions to the cells, are often deformable or viscoelastic, and expose cells to stimuli such as chemical gradients and fluid flow. This gap in knowledge hinders attempts to predict the spread of infections, design new bacterial therapies, develop accurate ecological models, and implement effective agricultural and bioremediation strategies. Elucidating how complexities such as physical confinement, environmental deformability and viscoelasticity, and external stimuli alter bacterial motility and spreading is therefore an important frontier of current research that bridges biology, physics, and engineering.

Moreover, bacteria are an attractive system to use in fundamental studies of active matter due to the diversity and tunability of their transport behavior. For example, the run-and-tumble dynamics exhibited by \textit{E. coli}, \textit{B. subtilis}, and many other bacteria are also observed in diverse other active systems, including self-propelled colloids, algae, eukaryotic cells, and animals ~\citep{karani2019tuning,lozano2018run,polin2009chlamydomonas,heuze2013migration,benichou2011intermittent}. As a result, a growing number of studies describe active particles using run-and-tumble dynamics, and results obtained with bacteria may be applicable to other systems as well. Furthermore, the use of different bacterial species and genetic mutants enables the study of cells of diverse other motility behaviors, shapes, sizes, and surface chemistries, which provides a means to explore a broad range of microswimmer properties ~\citep{tokarova2021patterns,harman,theves2013bacterial,Xie:2011,packer1997rhodobacter,kinosita2018unforeseen,kuhn2017bacteria,jin2011bacteria}. Thus, studies of bacterial transport in complex environments can provide useful fundamental insights into the dynamics of active matter more broadly.  

In this section, we review recent studies of the single- and multi-cellular transport of bacteria---focusing on rod-shaped run-and-tumble bacteria as a model system---in complex environments characterized by physical confinement, viscoelasticity, and forcing by external stimuli such as chemical gradients and fluid flow. Because our focus is on active transport, we only focus on the dynamics of planktonic, non surface-attached bacteria. However, studying the formation of surface-attached bacterial communities in complex environments is another active area of current research that has been reviewed recently in e.g., Refs.~\citenum{wong2021roadmap} and \citenum{conrad2018confined}. Furthermore, we note that many bacteria self-propel through other mechanisms, such as by swarming or pulling themselves on surfaces ~\citep{jin2011bacteria}---which potentially enables them to sense mechanical cues on surfaces ~\citep{Persat:2015} and thereby respond to a broader range of stimuli. Investigating these other forms of active transport in complex environments is another important area of current research ~\citep{tokarova2021patterns}. Finally, motivated by rapid progress in the synthesis of colloidal microswimmers capable of recapitulating many of the features of bacteria ~\citep{Bechinger2016b}, we also integrate references to studies of synthetic microswimmers throughout this section where relevant. Such systems enable swimmer shape, deformability, and propulsion mechanism to be tuned even more broadly, and investigating the influence of these factors on active transport will continue to be a useful direction for future work.

\subsection{Complex geometries and obstacles}\label{bac-geometry}
\begin{figure}
  \begin{center}
        \includegraphics[width=\columnwidth]{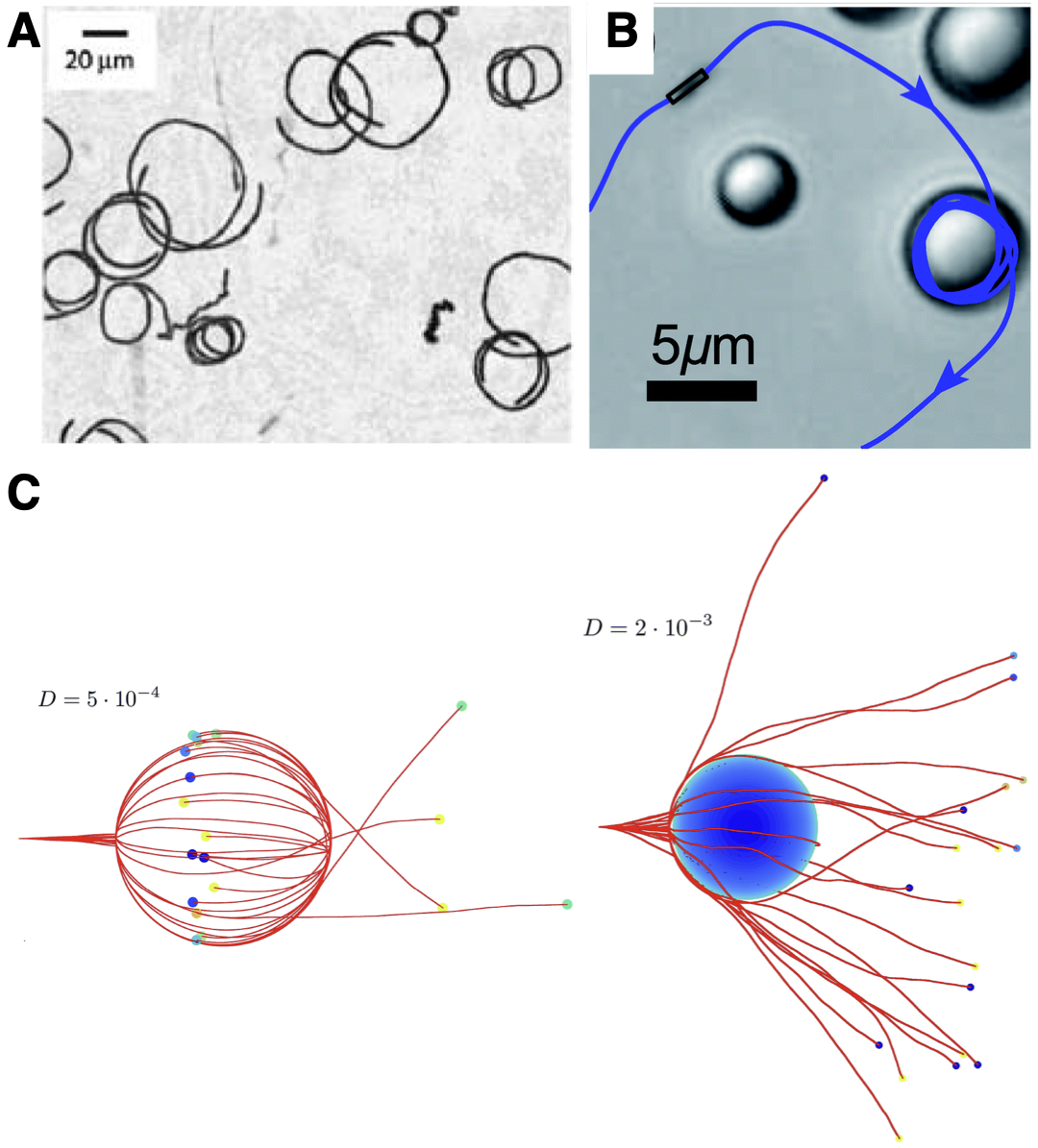}
    \caption{\begin{small}\textbf{Microswimmers moving near flat and curved surfaces.} \textbf{A}. Superposition of phase-contrast video microscopy images showing \textit{E. coli} cells swimming in circular trajectories near a flat glass surface. Reprinted with permission from Ref.~\citenum{Lauga:2006} (Elsevier, 2006). \textbf{B}. Trajectory of a chemically propelled $2~\mu$m-long micro-rod orbiting around a passive sphere of diameter $\sim6~\mu$m. Reprinted with permission from Ref.~\citenum{Takagi:2014}  (The Royal Society of Chemistry, 2014). \textbf{C}. Twenty instances of computed trajectories for a pusher microswimmer incident upon a large sphere for smaller (left) and larger (right) microswimmer translational diffusion constants $D$, which have been made dimensionless as detailed further in Ref.~\citenum{spagnolie2015geometric}. The left panel shows an example in which microswimmers remain hydrodynamically bound to the sphere surface; color coding/shading indicates the final position in the normal direction with darker swimmers coming out of the page and lighter swimmers going into the page. The right shows an example in which microswimmer motion is rectified, but they do not remain bound. Reprinted with permission from Ref.~\citenum{spagnolie2015geometric} (The Royal Society of Chemistry, 2015).\end{small}
    \label{fig:bact-figure1}}
  \end{center}
\end{figure}

\textbf{Swimming near flat and curved surfaces.} The simplest form of physical confinement is being forced to interact with a solid surface. In this case, hydrodynamic and physicochemical interactions retain cells near the surface ~\citep{Takagi:2014,Brown:2016,Frymier:1995,Lauga:2006,molaei,wensink2008aggregation,elgeti2009self,schaar2015detention,spagnolie2012hydrodynamics,berke2008hydrodynamic,vigeant2002reversible,li2009accumulation,sipos2015hydrodynamic}, although additional corrugations on the surface can suppress this effect ~\citep{kurzthaler2021microswimmers,mok2019geometric}. Similar behavior arises at an immiscible liquid interface ~\citep{vladescu2014filling,desai2018hydrodynamics,ahmadzadegan2019hydrodynamic}, although additional effects arising from interfacial flows can also play a role ~\citep{deng2020motile}. When the surface is flat, or its radius of curvature is much larger than the cell size, an \textit{E. coli} cell continues to swim in planar clockwise (when viewed from above) circles along the surface (Fig.~\ref{fig:bact-figure1}\textbf{A})---reflecting the clockwise (when viewed from behind the cell) counter-rotation of the cell body in response to the counter-clockwise rotation of the flagellar bundle that propels the cell ~\citep{Frymier:1995,Lauga:2006}. More generally, the direction of this circular motion presumably depends on which direction the flagella rotate; further details of the underlying physics are provided in the chapter \textit{Hydrodynamics of Cell Swimming}. Random fluctuations or tumbles then enable the cell to escape. Conversely, when the radius of curvature of the surface of an obstacle is comparable to or smaller than the cell size, the obstacle only rectifies the direction of swimming, causing the cell to be transiently retained at the surface before continuing to move beyond it ~\citep{spagnolie2015geometric,Takagi:2014} (Fig.~\ref{fig:bact-figure1}\textbf{B-C}).\\ 

\begin{figure*}[htp]
  \begin{center}
        \includegraphics[width=0.7\textwidth]{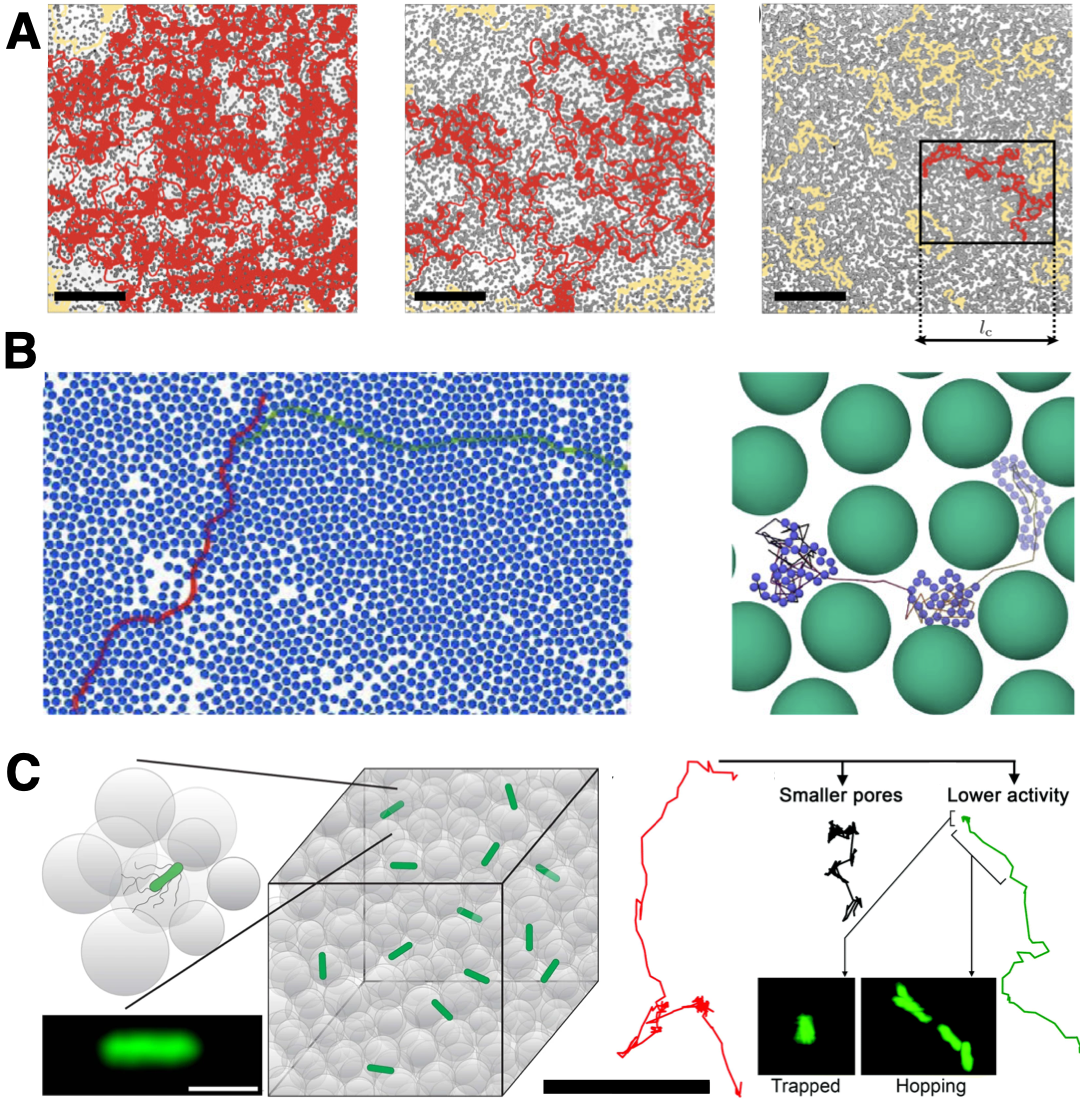}
    \caption{
\begin{small}\textbf{Hindered motility of microswimmers in crowded environments.} \textbf{A}. Colloidal rollers in random obstacle lattices of different densities. Trajectories of colloidal rollers (red and yellow) are superimposed on the pictures of the obstacle lattices. Scale bar is $500~\mu$m. Obstacle density increases from left to right. At the lowest density, the trajectories form a single percolating cluster; at the intermediate density, the trajectories form disconnected clusters (yellow), while the largest cluster (in red) still percolates through the observation region; at the highest obstacle density, none of the disconnected clusters percolate, and no macroscopic transport is observed. The largest cluster of maximal dimension $l_c$ is colored in red. Reprinted with permission from Ref.~\citenum{morin} (American Physical Society, 2017). \textbf{B}. Left shows the computed trajectory of a stiff, elongated, active filament in a random obstacle lattice; right shows the computed trajectory of a flexible active filament, which curls up and intermittently jumps to nearby sites. Reprinted with permission from Ref.~\citenum{Mokhtari:2019} (American Physical Society, 2019). \textbf{C}. Left shows a schematic of a transparent, 3D porous medium comprised of a jammed packing of hydrogel particles swollen in liquid cell culture medium (gray circles). Fluorescent \textit{E. coli} (green rods) swim through the pores between particles; inset shows a micrograph of a GFP-labeled bacterium. Scale bar represents $2~\mu$m. Right shows trajectories of cells moving through the pore space \textit{via} hopping-and-trapping motility; in a medium with smaller pores, hopping lengths are smaller and trapping durations are longer, while for a cell with lower activity, hopping lengths are unchanged but trapping durations are longer. Inset images show selected superimposed frames of cellular motion, equally spaced in time, showing that the cell is localized and randomly oriented during trapping, but moving and directed during hopping. Scale bar is $20~\mu$m. Adapted with permission from Refs.~\citenum{tapa19a} and \citenum{tapa19b} (Springer Nature and The Royal Society of Chemistry, 2019).\label{fig:bact-figure2}\end{small}}
  \end{center}
\end{figure*}

\noindent \textbf{Transport in structured 2D environments.} This phenomenon can have important consequences for individual bacteria moving in crowded and porous media, in which they are forced to interact with many surrounding solid surfaces. Most three-dimensional (3D) porous media are opaque, typically precluding direct observation of bacteria and other microswimmers within the pore space. MRI measurements provide quantification of macroscopic spreading ~\citep{sherwood2003analysis}; however, they do not have sufficient sensitivity to probe single-cell behavior. As a result, experiments typically focus on simplified, two-dimensional (2D) models of porous media made by planar confinement of arrays of cylindrical obstacles. These studies, corroborated and extended by simulations, have catalogued a diversity of transport behaviors. Intuitively, one expects that the presence of obstacles hinders transport---which is indeed the case for microswimmers in disordered media with densely-packed obstacles ~\citep{Mokhtari:2019,morin,sosa2017motility,raatz2015swimming,volpe2011microswimmers,martinez2018collective,martinez2020trapping}, even causing surface-mediated localization in sufficiently dense media ~\citep{morin,hofling2013anomalous,van1982transport,Bertrand:2018,zeitz2017active,jakuszeit2019diffusion,chamolly2017active,aragones2018diffusion,paoluzzi2014run,guccione2017diffusivity,Reichhardt:2014,reichhardt2018,martinez2020trapping} (Fig.~\ref{fig:bact-figure2}\textbf{A-B}). In ordered media, however, surface interactions can help guide swimmers through the straight interstitial channels of the obstacle lattice, promoting size-dependent swimmer transport ~\citep{Brown:2016,dietrich2018active,chopra2020geometric}. Moreover, at low densities of small obstacles, bacterial transport can be unexpectedly \textit{enhanced} beyond the case of obstacle-free environments due to rectification of the circular planar motion of the cells as they encounter obstacles ~\citep{Makarchuk:2019}.\\

\noindent \textbf{Transport in structured 3D environments.} While these 2D studies provide important intuition, how such effects translate to 3D porous media is unclear. For example, the pore space of a 3D medium is more connected, and the threshold solid fraction at which pores can percolate through the medium is lower than in a 2D medium ~\citep{sahini2003applications}. As a result, bacteria have more paths and degrees of freedom available to them in 3D. In addition, the flow field generated by a microswimmer is fundamentally different in 3D compared to 2D media ~\citep{lauga2009hydrodynamics}. Simulations provide clues that these effects strongly impact microswimmer behavior, suggesting that active transport in 2D and 3D media may be fundamentally different ~\citep{morin,stenhammar2014phase}. Thus, there is growing interest in experiments capable of probing active transport in 3D porous media. 

One approach is to study cells inoculated in viscoelastic ``semisolid'' agar---a common microbiology assay for bacterial motility ~\citep{tittsler}. When the cells are sufficiently dense, their macroscopic spreading through the porous 3D agar matrix can be visualized using a conventional camera. In a seminal study, Wolfe and Berg~~\citep{Wolfe:1989} used this approach to study the spreading of mutants of \textit{E. coli}, including those whose tumbling rate could be controlled chemically. Intriguingly, they found that the extent of spreading in the agar matrix was non-monotonic with the cell tumbling rate: while incessantly tumbling cells did not appreciably spread, as one may expect, smooth swimming cells that do not tumble at all also did not appreciably spread. Instead, these cells became entrapped in the agar matrix---revealing that tumbling is essential for cells to spread in complex, porous environments. Theory and simulations also capture this non-monotonic behavior and have demonstrated it more generally for diverse microswimmers in media with different obstacle geometries ~\citep{licata,Bertrand:2018,Volpe:2017,christinageometric2021}. Related work has also shown that bacterial spreading is similarly non-monotonic in the number of flagella used by the cells, which tunes the amount by which their cell bodies reorient during tumbles ~\citep{Najafieaar6425}. Thus, bacterial spreading in crowded and porous environments likely depends on the frequency and amplitude of reorientations as well as the geometry of the environment, which influences how cells move and collide with surrounding obstacles. Further exploration of this interplay, both for bacteria and other microswimmers more broadly, will be an interesting direction for future research.

While experiments using semisolid agar provides powerful insights into the ability of cells to macroscopically spread, they are limited in their ability to elucidate single-cell behavior; such media are turbid, precluding high-fidelity and long-time visualization of individual cells. Moreover, there is limited control over the properties of the agar matrix, and thus, such media often do not have well-defined pore structures. Hence, to overcome these limitations, researchers have developed 3D porous media made of densely-packed hydrogel particles swollen in defined liquids containing the salts and nutrients required to sustain cellular functioning ~\citep{tapa19b} (Fig.~\ref{fig:bact-figure2}\textbf{C}, left). The individual cells cannot penetrate into the particles; instead, cells move by swimming through the pores formed between adjacent hydrogel particles. The stress exerted by cellular swimming is insufficient to deform the solid matrix, and thus, the packings act as rigid, static matrices. Tuning the hydrogel particle packing density provides a straightforward way to tune the pore size distribution of the overall packing, and thereby modulate cellular confinement. Moreover, because the particles are highly swollen, these packings are transparent, enabling cellular transport to be interrogated \textit{in situ} at single cell resolution using confocal microscopy.


Experiments using this platform have revealed fundamental differences in the transport of \textit{E. coli} in 3D porous media compared to in bulk liquid ~\citep{tapa19a,tapa19b}. In particular, analysis of the trajectories of single cells demonstrated that the paradigm of run-and-tumble motility does not apply to cells in tight porous media with mean pore sizes between $\sim1$ to $10~\mu$m---characteristic of many bacterial habitats ~\citep{dullien,fatin,lang,zalc,lindquist}, and comparable to the overall size of a single cell. Instead, the cells exhibit a distinct mode of motility known as ``hopping-and-trapping'' (Fig.~\ref{fig:bact-figure2}\textbf{C}, right). To move through the pore space, a cell must perform a run along a straight line path; however, it collides with an obstacle well before it completes its run. Thus, runs are truncated (termed ``hops''), with the hopping lengths set by the lengths of the straight line paths that fit in the pore space (known as ``chords'' ~\citep{torquato,lu92}). Collisions with the surrounding matrix do not act as tumbles that rapidly reorient the cell, as is often assumed; instead, steric interactions between flagella and the surrounding solid matrix cause the cell to become transiently trapped, as first observed by Wolfe and Berg ~\citep{Wolfe:1989} and corroborated by experiments in 2D microfluidic models as well ~\citep{tokarova2021patterns}. When a cell is trapped, its flagella remain bundled and continue to rotate---in some cases, over durations much longer than the duration of runs in unconfined liquid, indicating that confinement suppresses unbundling ~\citep{tapa19a}. The torque generated causes the cell body to constantly reorient until the flagella have room to unbundle and rebundle on the opposing pole of the cell body, enabling the cell to escape---similar to the swim-pause-reverse behavior seen for bacteria in model wedge-shaped geometries ~\citep{Cisneros:2006}. After it escapes, the cell then continues to move through the pore space along another hop until it again encounters an obstacle and becomes trapped. Thus, trapping of cells is an active matter analog of the intermittent trapping of passive species that arises during thermally-activated transport in diverse disordered systems ~\citep{bouchaud}, such as colloidal particles in dense suspensions or polymer networks, adsorbing solutes in porous media, macromolecules inside cells, molecules at membranes, and even charges in amorphous electronic materials ~\citep{scher,weeks02,wong,drazer,akimoto,yamamoto13,yamamoto14}.

Indeed, by analogy to the process by which large polymers thermally escape from tight pores in a disordered porous medium ~\citep{muthu87,muthu89a,muthu89b}, tight spots of a crowded 3D environment can be thought of as entropic traps for cells ~\citep{tapa19b}, within which only certain configurations of the cell enable it to escape ~\citep{han}. The duration of trapping is then determined by a competition between cellular activity ~\citep{Woillez2019,Geiseler2016,Bi2016,takatori2015towards} and confinement (Fig.~\ref{fig:bact-figure2}\textbf{C}, right). Unlike conventional tumbles, trapping lasts $\tau_{\rm{t}}\sim1$ to $10$ s, longer than the hopping duration $\tau_{\rm{h}}\sim0.1$ to $1$ s. Hence, over large length and time scales, each cell performs a random walk with pauses between successive steps, reminiscent of a L\'{e}vy walk with rests ~\citep{Zaburdaev:2015}---suggesting that bacteria swimming in a porous medium have unexpected similarities to moving mammalian cells ~\citep{harris2012generalized}, robots searching for a target ~\citep{nurzaman2009yuragi}, and tracers in a chaotic flow ~\citep{weeks1997experimental}. This process can therefore be modeled using a stochastic two-state random walk approach ~\citep{lazaro}. The characteristic step length is given by the mean hopping length, $\bar{\ell}_{\rm{h}}\sim1$ to $10~\mu$m, which is set by the geometry of the porous medium, and the characteristic step time is primarily determined by the mean trapping duration, $\bar{\tau}_{\rm{t}}\sim1$ to $10$ s, which is set by both medium geometry and cellular activity. These quantities then yield a translational diffusion coefficient $D_{\rm{b}}\sim\bar{\ell}_{\rm{h}}^{2}/\bar{\tau}_{\rm{t}}\sim0.1$ to $100~\mu$m$^{2}$/s, much smaller than the unconfined case, that can quantitatively describe bacterial spreading in 3D porous media over large length and time scales ~\citep{tapa19a}. \\

 \begin{figure}
        \includegraphics[width=\columnwidth]{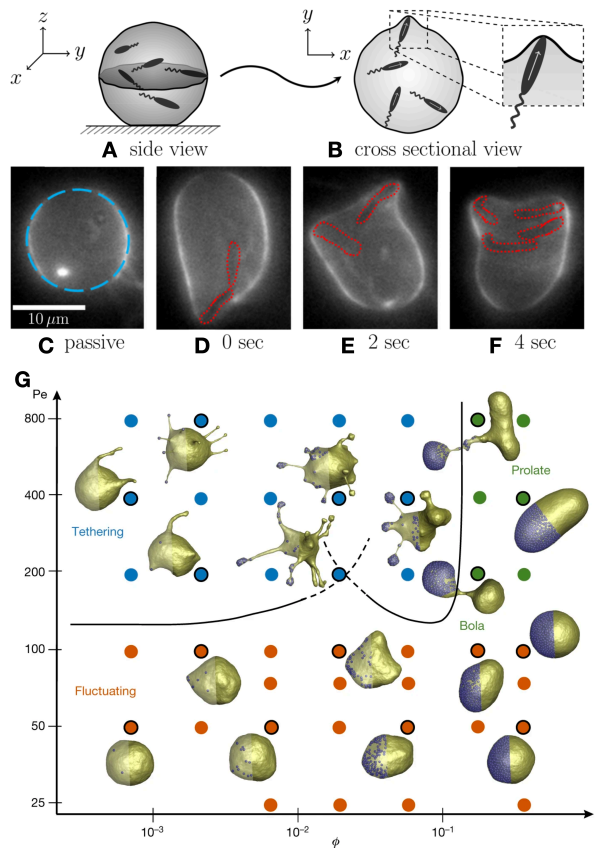}
    \caption{\begin{small}\textbf{An example of active transport in a deformable medium.} Giant unilamellar vesicle containing motile \textit{Bacillus subtilis}. The schematic shows how the 3D system (A) is imaged at a single equatorial cross section (B) to generate the experimental images in (C)-(F). The dashed blue outline in (C) shows the undeformed spherical shape of the membrane when bacteria are non-motile, while (D)–(F) show how motile bacteria (dotted red outlines) generate large membrane deformations at different times. Reprinted with permission from Ref.~\citenum{PhysRevLett.124.158102} (American Physical Society, 2020). (G) shows results from simulations of self-propelled particles enclosed by a deformable membrane. This diagram maps the different states that arise for different particle P\'{e}clet numbers (Pe), which indirectly quantifies the propulsion force, and volume fractions $\phi$. Three regimes emerge: tethering (blue symbols), fluctuating (red symbols) and bola/prolate (green symbols) vesicle shapes. The points corresponding to the snapshots have black outlines. Further details are provided in the chapter \textit{Computational Physics of Active Matter}. Reprinted with permission from Ref.~\citenum{vutukuri2020active} (Springer Nature, 2020).\end{small}\label{fig:bact-figure3}}
\end{figure}

\noindent \textbf{Additional open questions.} While these studies provide a revised view of motility in crowded and porous environments, and draw intriguing connections between out-of-equilibrium, active matter and passive systems, they also motivate many fascinating questions to be addressed by future work. For example: How do the characteristics of hopping and trapping motility depend on cellular morphology and surface properties, as well as the geometry of the surrounding crowded environment? How do these behaviors change for other forms of active matter that employ other motility behaviors in unconfined environments distinct from run-and-tumble dynamics? How do chemical interactions (e.g., surface adhesion) and mechanical interactions (e.g., deformations of cells as well as their surroundings), which arise in many complex environments, influence active transport? One example of how mechanical interactions, in which cellular swimming deforms a surrounding boundary, give rise to intriguing dynamics is shown in Fig. \ref{fig:bact-figure3}; other examples are described in Section \ref{bac-polymer} below.

Furthermore, how do these behaviors change for more concentrated bacterial populations? In bulk fluid, short-range interactions between active particles, such as bacteria, enable them to coordinate motion and spontaneously ``flock'' along the same direction or coherently ``swirl'' ~\citep{Driscoll2017a,martin2018collective,ginelli2010large,dunkel2013fluid,dombrowski2004self,cisneros2010fluid,ishikawa2011energy,wensink2012meso,colin2019chemotactic}. At higher concentrations, they can even form dense clusters or completely phase separate, depending on the interactions between them ~\citep{takatori2015towards,cates2015motility,theurkauff2012dynamic,buttinoni2013dynamical,pohl2014dynamic,palacci2013living,aragones2019aggregation,chen2015dynamic,petroff2015fast,ginot2015nonequilibrium,redner2013reentrant,fily2012athermal,digregorio2018full,thutupalli2018flow,geyer2019freezing}. When the cells are globally confined (e.g., in a thin cylindrical chamber), such collective behaviors can be promoted \cite{wioland2013confinement,lushi2014fluid} and potentially even harnessed to drive larger-scale autonomous motion \cite{gao2017self,young2021many,sanchez2012spontaneous}. However, when cells are surrounded by a disordered array of obstacles, this solid matrix suppresses these short-range interactions and thus can suppress such collective behaviors ~\citep{martin2018collective,Chepizko:2013,altenberger1986hydrodynamic,chepizhko2015active,chepizhko2013optimal,sandor2017dynamic,sandor2017dewetting,sese2018velocity}. Nevertheless, other forms of coordination that rely on the coupling of cellular motion to external stimuli that can extend over longer distances, such as chemical fields, can still persist; these are described further in Section \ref{bac-stimuli}.







\subsection{Viscoelastic polymer solutions}\label{bac-polymer}
Many bacterial habitats are polymeric fluids. Prominent examples include mucus in the lungs and digestive tract ~\citep{lai2009micro,celli2009helicobacter,steinberg2019high,pnas}, exopolymers in the ocean ~\citep{decho1990microbial}, and cell-secreted extracellular polymeric substances (EPS) that encapsulate surface-attached biofilms ~\citep{dorken2012aggregation}. One may expect that, for a given bacterial motor torque, the increased fluid viscosity caused by polymers would lead to slower cellular swimming. Surprisingly, however, experiments show that increasing polymer concentration increases swimming speed, sometimes followed by a decrease in speed at large concentrations ~\citep{schneider1974effect,berg1979movement}. In some cases, this behavior may reflect the metabolism of polymers by bacteria; however, it is also observed with non-metabolizable polymers, suggesting a physicochemical role played by polymers as well. While fully elucidating the relative influence of fluid viscosity, viscoelasticity, and shear thinning on this behavior is still an outstanding challenge, recent work has started to shed light on the physics of bacterial swimming in polymer solutions. \\

 \begin{figure}[htp]
  \begin{center}
        \includegraphics[width=\columnwidth]{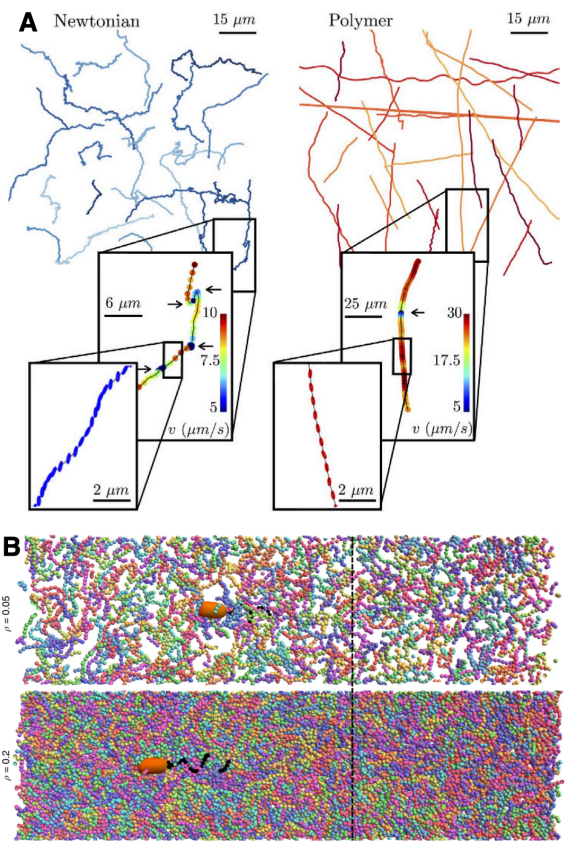}
    \caption{\begin{small}\textbf{Bacterial swimming in viscoelastic polymer solutions.} \textbf{A}. Kinematics of swimming \textit{E. coli} cells in both Newtonian and viscoelastic fluids. Left and right panels show trajectories of cells in Newtonian buffer and in viscoelastic carboxy-methyl cellulose solution. Cells in polymer solution move remarkably straighter compared to cells in buffer. Sample cell trajectories in the magnified view show run-and-tumble motility i.e. nearly straight lines connected at random angles (tumbles denoted by arrows). Further magnified views show that the cell body orientation  oscillates or `wobbles' in the buffer solution, but wobbles diminish in the polymer solution. Reprinted with permission from Ref.~\citenum{patteson2015running} (Springer Nature, 2015). \textbf{B}. Typical simulation snapshots of bacterial dynamics in fluids consisting of a Newtonian background fluid and including semiflexible polymers at low (top) and high (bottom) volume fractions. The starting position of the cell body is indicated by the black dashed line, showing that the bacterium swims faster in the more concentrated suspension. The colours of individual polymers are to aid visualization. Reprinted with permission from Ref.~\citenum{zottl2019enhanced} (Springer Nature, 2019).\end{small}\label{fig:bact-figure4}}
  \end{center}
\end{figure}

\noindent\textbf{Transport in dilute polymer solutions.} In dilute solutions of long and flexible polymers, \textit{E. coli} exhibit three notable changes in their swimming kinematics ~\citep{patteson2015running} (Fig. \ref{fig:bact-figure4}\textbf{A}): (i) the cells tumble less frequently, (ii) the cells ``wobble" less during runs, and (iii) the swimming speed is increased. By tracking cells in solutions of polymers of different molecular weights, viscosities, and fluid elasticities, the authors of Ref. \citenum{patteson2015running} proposed that the first effect reflects the increased viscosity of the solution due to polymers; as viscous stresses increase, the mechanical load on the flagellar motor increases, which increases the effective energy barrier for the cell motor to switch and thereby decreases the tumbling rate ~\citep{qu2018changes,fahrner2003bacterial,yuan2009switching}. By contrast, the second and third effects are instead thought to reflect the increased elasticity of the solution due to polymers. In polymer-free solutions, individual cells appear to wobble during their runs, due to the projection of the cell's helical trajectory in three dimensions that is produced by the counter-rotating cell body and flagellar bundle ~\citep{watari2010hydrodynamics}. The curvature associated with this helical trajectory stretches surrounding polymers ~\citep{pakdel1996elastic}, producing elastic hoop stresses that stabilize wobbling and direct more of the thrust in the forward direction, enabling the cell to swim faster ~\citep{PhysRevFluids.6.053301,housiadas2021squirmers,binagia2020swimming}. Because these effects all depend closely on the specific mechanics of flagellar propulsion, understanding how they manifest for other microswimmers that employ other methods of self-propulsion is another area of active inquiry ~\citep{zhu2012self,qin2015flagellar}. Indeed, analysis of the fluid dynamics of diverse other microswimmers indicates that elastic stresses can induce faster swimming even for swimmers that do not wobble \cite{liu2011force,castillo2019drag,pak2014generalized,puente2019viscoelastic,rogowski2021symmetry,binagia2020swimming}. 

Elastic stresses may not be the only---or even the dominant---cause, however, of observations (i)--(iii) noted above; indeed, recent experiments studying \textit{E. coli} swimming in colloidal suspensions reported similar behavior \cite{kamdar2021}. Thus, the authors suggested that the hydrodynamic interaction between individual bacteria and the colloidal component of complex fluids in general, mediated by the background Newtonian fluid, is instead the primary cause of these swimming kinematics. Further work disentangling the relative influence of elastic stresses and other hydrodynamic interactions on active transport will therefore be a useful direction for future research.\\

\noindent \textbf{Transport in more concentrated polymer solutions.} When the solution is semidilute, interactions between polymer molecules play an appreciable role as well. Physical entanglements or chemical crosslinks can cause the polymers to form a rigid network that obstructs cells, leading to effects such as those described in Section \ref{bac-geometry}. In the absence of such network formation, or over time scales long enough for the network to relax, the solution rheology and microstructure regulate cellular swimming. At the scale of a single cell, the fast-rotating flagellar bundle generates strong shear locally near the `tail' of the cell body, causing neighboring polymer molecules to align and locally reduce the solution viscosity. For a given motor torque, this viscosity reduction at the flagella is thought to enable the cell to swim faster ~\citep{martinez2014flagellated,qu2020effects,zhang2018reduced}. Simulations also suggest that the polymer molecules themselves also distribute non-uniformly in the vicinity of the bacterium, with polymer being depleted inside and around the flagella (Fig. \ref{fig:bact-figure4}\textbf{B}). This depletion leads to an apparent slip that, again, is thought to also enable the cell to swim faster ~\citep{zottl2019enhanced}. \\

\begin{figure}[htp]
  \begin{center}
        \includegraphics[width=\columnwidth]{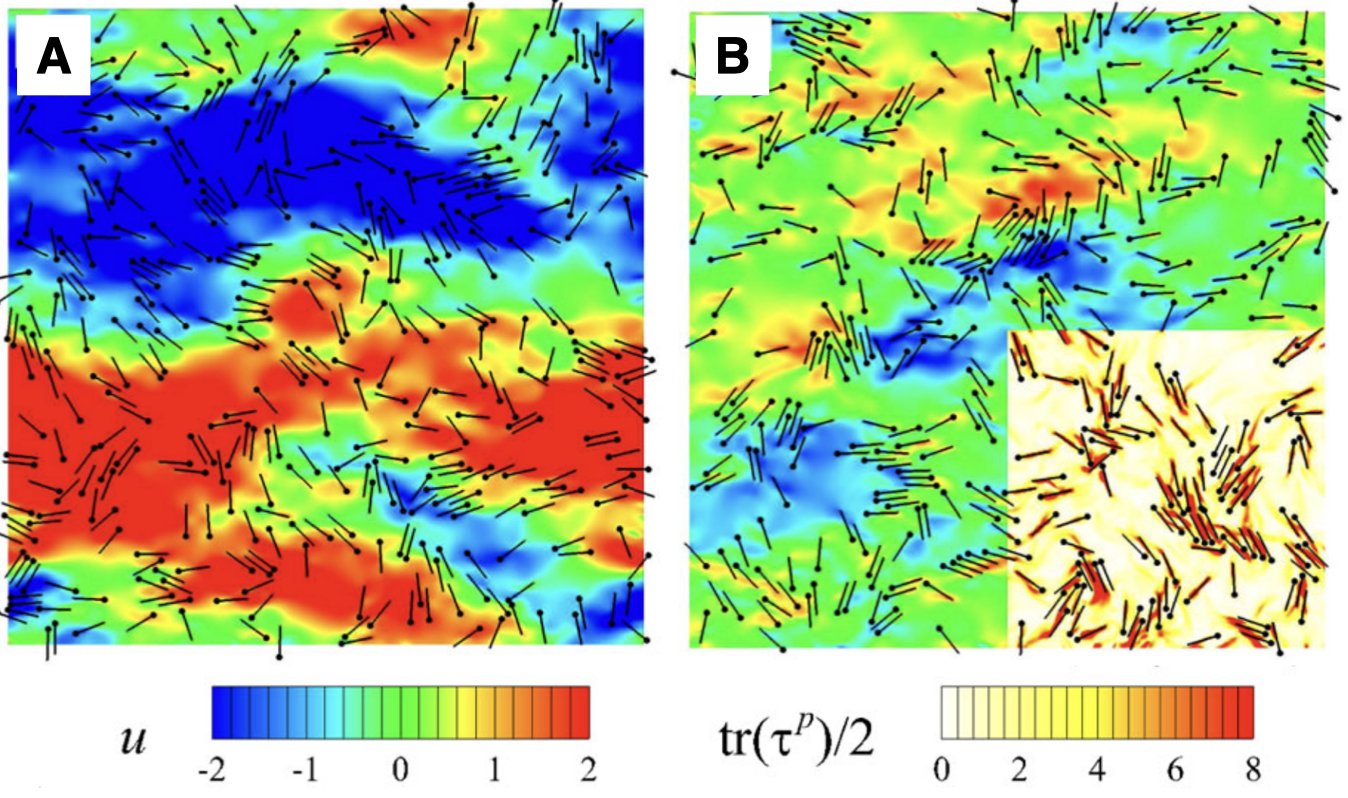}
    \caption{\begin{small}\textbf{Viscoelasticity can suppress coherent motion in concentrated bacterial populations.} Computed flow field and distribution of `pusher' microswimmers in a \textbf{A} Newtonian and \textbf{B} viscoelastic fluid. Contours in the main frame and the right-bottom quarter of \textbf{B} show the dimensionless horizontal velocity component $u$ and the dimensionless elastic energy given by the trace of the polymer contribution to the stress tensor tr$(\tau^p)/2$, respectively. Reprinted with permission from Ref.~\citenum{li2016collective} (American Physical Society, 2016).\end{small}\label{fig:bact-figure5}}
  \end{center}
\end{figure}

\noindent \textbf{Additional behaviors that arise in polymeric environments.} How do these behaviors change for more concentrated bacterial populations? As noted in Section \ref{bac-geometry}, in bulk fluid, short-range interactions enable bacteria and other forms of active matter to coordinate their motion over large scales. In a polymeric fluid, the elastic stresses generated by swimming are thought to drive bacteria together, enhancing their local aggregation and, in some cases, suppressing large-scale coherent motions and morphological instabilities ~\citep{li2016collective,emmanuel2020active} (Fig. \ref{fig:bact-figure5}). However, when concentrated suspensions are physically confined as well, polymer stresses can conversely \textit{promote} coherent motion over length scales much larger than in bulk fluids---for example, driving the formation of uniformly rotating `giant vortices' ~\citep{liu2021viscoelastic}. Another effect arises when cells are sufficiently close to each other; in a polymer solution, if two cells are closer than the polymer size, they deplete the polymer from the space in between them. As a result, the unbalanced osmotic pressure of the solution forces the cells together. This attractive depletion interaction can drive the aggregation of both non-motile and motile cells in a manner similar to passive colloidal particles ~\citep{dorken2012aggregation,schwarz2010polymer}, although the swimming stress generated by motile cells suppresses aggregation ~\citep{porter2019interplay}. Intriguingly, for the case of motile cells, the finite size aggregates that form \textit{via} depletion interactions show unidirectional rotation, driven by the torques exerted by the bacteria at the outer periphery of each aggregate ~\citep{schwarz2012phase}. Disentangling the influence of these different behaviors, as well as other behaviors that can arise for microswimmers with other swimming characteristics \cite{elfring2010two,elfring2015theory,datt2015squirming,datt2017active,pietrzyk2019flow}, will be an important direction for future research.

\subsection{External stimuli}\label{bac-stimuli}
\textbf{Transport in response to chemical stimuli.} An important feature of many bacteria, as well as other forms of active matter ~\citep{saha2014clusters,liebchen2019interactions,Liebchen2018a,popescu2018chemotaxis,singh2019competing,marsden2014chemotactic,agudo2019active,Illien2017,Stark2018}, is their ability to sense chemical stimuli and respond by modulating motility. As the cells swim in bulk liquid, they sense varying concentrations of different chemicals---e.g., sugars, amino acids, and oxygen---that continually bind and unbind to and from cell-surface receptor complexes. This binding triggers a network of subcellular processes that primarily modulate the frequency of tumbling and bias run length. Cells thus perform longer or shorter runs when moving toward regions of higher or lower concentration of chemoattractants, respectively, and exhibit the opposite behavior in their response to chemorepellents. As a result, the random walk determined by run-and-tumble motion is biased ~\citep{berg}; the cell's intrinsic ability to bias its motion is described by a parameter, having the same units as the diffusion coefficient, that is known as the chemotactic coefficient $\chi$ ~\citep{lauffenburger1991quantitative,Keller1971a}. The flux due to chemotaxis is then given by $bv_{c}$, where $b$ is the number density of cells and $v_{c}\equiv\chi\boldsymbol{\nabla} f(c)$ is known as the chemotactic velocity; the function $f(c)$ describes the ability of the bacteria to logarithmically sense nutrient at a concentration $c$ ~\citep{Cremer2019,fu2018spatial,dufour,yang2015relation}.

\begin{figure*}[htp]
  \begin{center}
        \includegraphics[width=0.7\textwidth]{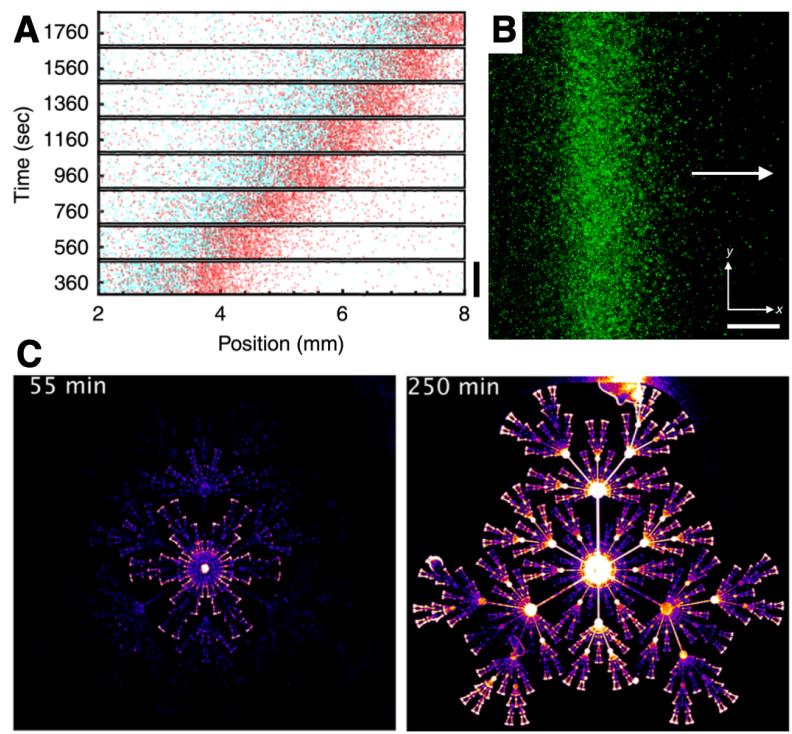}
    \caption{\begin{small}\textbf{Chemotactic migration of bacterial populations.} \textbf{A}. Images taken at different time points (vertical axis) showing \textit{E. coli} performing chemotactic migration in a straight microfluidic channel. Red and cyan dots show cells with low and high tumble bias---the fraction of time a cell spends tumbling---indicating that cells self-organize within a traveling band. Scale bar shows 0.6 mm. Reprinted with permission from Ref.~\citenum{fu2018spatial} (Springer Nature, 2018). \textbf{B}. Magnified
bottom-up fluorescence intensity projection of a chemotactic front propagating within a 3D porous medium, showing individual cells of \textit{E. coli}. Arrow indicates direction of overall propagation. Scale bar denotes $200~\mu$m. Reprinted with permission from Ref.~\citenum{tapabiophys} (Elsevier, 2021). \textbf{C}. Microscopy images showing the outward spreading of chemotactic bacteria in 2D fractal microfluidics; time stamp indicates time elapsed after initial inoculation in the center. Reprinted with permission from Ref.~\citenum{PhysRevX.10.031017} (American Physical Society, 2020).\end{small}\label{fig:bact-figure6}}
  \end{center}
\end{figure*}

At the multicellular level, this process of chemotaxis can mediate directed collective motion, or migration. A particularly striking example arises when the cells continually consume a surrounding attractant, such as a nutrient or oxygen: the cells collectively generate a local gradient that they in turn bias their motion along, leading to the formation of a coherent band of cells that continually propagates ~\citep{adler1966science,adler1966effect,Cremer2019,fu2018spatial,saragosti2011directional,douarche2009coli,tuval2005bacterial}, sustained by its continued consumption of the surrounding attractant. This phenomenon was first demonstrated by Adler in 1966 through an elegant experiment in which he placed a dense plug of \textit{E. coli} at one end of a capillary tube containing nutrient-rich liquid ~\citep{adler1966science}; the cells spectacularly migrated through the tube as bands visible to the naked eye. Since then, continuum-scale models coupling attractant dynamics to bacterial fluxes have been developed ~\citep{lauffenburger1991quantitative,Keller1971a} and can successfully capture the key features of chemotactic migration in bulk liquid ~\citep{Keller1971a,fu2018spatial,Seyrich2019} (Fig. \ref{fig:bact-figure6}\textbf{A}) and in viscoelastic agar ~\citep{croze,Cremer2019}. Additional studies using a range of different chemoattractants have shown that this phenomenon of chemotactic migration can enable populations to escape from harmful environments or to colonize new terrain ~\citep{Cremer2019,PhysRevX.10.031017}. 

Moreover, because this form of collective migration relies on the coupling between a population-generated chemical gradient---which can extend over long distances spanning hundreds of cells---and biased cellular motion along this gradient, different bacteria can collectively influence and coordinate each other's motion across long distances, solely through chemoattractant consumption. This principle was recently demonstrated using 3D-printed cylindrical regions of \textit{E. coli} separated by different distances ~\citep{tapabiophys}. When the separation between the two cylinders was smaller than the length scale $\approx500~\mu$m over which nutrient is depleted by consumption, the cells were able to ``smell'' each other; as a result, chemotactic bands did not propagate between the two cylinders. However, when the separation between the two cylinders was much larger, cells migrated in chemotactic bands both outward and towards each other, in between the two cylinders. This observation suggests that, simply by tuning the geometry of a bacterial population, chemotactic migration can be directed and controlled.\\

\noindent \textbf{Chemotaxis in structured environments.} Chemotaxis can also enable collective migration to persist after a population is confronted with perturbations---either external ~\citep{sandor2017dynamic,Morin2017,Wong2014,Chepizhko2013,chepizhko2013optimal,chepizhko2015active,Toner2018,Maitra2020a}, stemming from heterogeneities in the environment, or internal, stemming from differences in the behavior of individual cells ~\citep{Yllanes2017,Bera2020,Alirezaeizanjani2020}. Such perturbations are usually thought to \textit{disrupt} collective migration. For example, as described in Section \ref{bac-geometry}, collisions with surrounding obstacles disrupt short-range coordination between cells ~\citep{sandor2017dynamic,Morin2017,Yllanes2017,Bera2020,Chepizhko2013,chepizhko2013optimal,chepizhko2015active,Toner2018}. However, as revealed through experiments on \textit{E. coli}, by generating large-scale nutrient gradients through consumption, bacterial populations can drive chemotactic migration, even in highly-confining, disordered, porous environments ~\citep{tapabiophys} (Fig. \ref{fig:bact-figure6}\textbf{B}). Nevertheless, several key differences arise in this case compared to that of migration in bulk liquid. First, the motility parameters $D_{\rm{b}}$ and $\chi$ are strongly reduced in a confinement-dependent manner, reflecting the hindered motion of individual cells in the tight pore space, as well as the increased occurrence of cell-cell collisions in the tight pore space---fundamentally altering the dynamics and morphology of a spreading population, as further explored theoretically in Ref. \citenum{amchin2021confinement}. Second, the cells use a different primary mechanism to perform chemotaxis. Instead of modulating the frequency of reorientations, which is not possible in tight environments, the cells primarily modulate the amount by which their cell bodies reorient by modulating the number of flagella that become unbundled during reorientations---thereby biasing their subsequent motion toward regions of higher nutrient concentration ~\citep{Berg:1972,vladimirov2010predicted,saragosti2011directional}. Thus, by employing multiple different mechanisms to bias their motion, cells can direct their transport in a range of different environments---motivating further studies of active matter transport in a variety of complex settings. Another recent example demonstrating how chemotaxis enables bacteria to escape from fractal mazes is shown in Fig. \ref{fig:bact-figure6}\textbf{C}.\\

\noindent\textbf{Morphological consequences of chemotaxis.} Perturbations are also typically thought to produce large-scale disruptions to velocity correlations between cells and the overall morphology of a cellular population ~\citep{Wong2014,Alert2020,Alert2019,Driscoll2017a,Doostmohammadi2016,williamson2018stability,saverio,bonilla2019contrarian} due to hydrodynamic ~\citep{koch,kela1,kela2} or chemical ~\citep{BenAmar2016,BenAmar2016b,funaki,brenner,mimura,stark} effects. However, recent calculations for suspensions of self-propelled particles have suggested that chemotaxis can stabilize such hydrodynamic instabilities ~\citep{Nejad2019}. Furthermore, experiments using \textit{E. coli} in straight channels have revealed that, even when cells have substantial differences in individual chemotactic abilities, they are able to migrate together with a common drift velocity by spontaneously sorting themselves to different positions with different local nutrient gradients (Fig. \ref{fig:bact-figure6}\textbf{B})---resulting in the formation of a chemotactic band that continues to migrate coherently ~\citep{fu2018spatial}. 

Other work studying \textit{E. coli} in 3D environments has shown that chemotaxis enables large-scale perturbations to the overall morphology of a population to be smoothed out as cells migrate, enabling them to continue to migrate together as a coherent band ~\citep{smoothing}. This population-scale smoothing unexpectedly reflects the manner in which individual cells transduce external signals during chemotaxis. In particular, it arises from limitations in the ability of cells to sense nutrient at high concentrations, resulting in spatial variations in cellular response to the local nutrient gradient. This behavior can be quantitatively described by spatial variations in the chemotactic velocity $\textbf{v}_{c}\propto\boldsymbol{\nabla} f(c)=f'(c)\boldsymbol{\nabla} c$; as in linear response theory, this velocity can be viewed as the bacterial response to the driving force given by the nutrient gradient, $\boldsymbol{\nabla}c$, modulated by the chemotactic response function $\chi f'(c)$. Thus, there are two distinct factors that influence how quickly cells move at different locations of a chemotactic band: the magnitude of the local nutrient gradient, and the magnitude of the local nutrient concentration itself, which sets the magnitude of the cellular response. Importantly, because the sensing function $f(c)$ is bounded, the response $f'(c)$ decreases at high nutrient concentrations, at which the cell surface receptors become increasingly saturated. As a result, regions of the population that are exposed to \textit{either} a weaker nutrient gradient $\boldsymbol{\nabla}c$ or more nutrient $c$, which leads to a weaker response $f'(c)$, can move slower than other regions. In the experiments of Ref. \citenum{smoothing}, outward-protruding regions of the population were exposed to more nutrient, causing the cells to respond more weakly and migrate slower, enabling the other regions to catch up and ultimately smoothing out morphological perturbations. Because migration in diverse other active matter systems also involves some form of limited sensing---e.g., chemotaxis by other bacteria ~\citep{menolascina2017logarithmic}, enzymes ~\citep{granick,Agudo-Canalejo2018a,Mohajerani2018}, amoeba ~\citep{keller1970initiation}, and mammalian cells ~\citep{Camley2018,iglesias,theveneau2010collective,mainikulesa,Malet-Engra2015,Puliafito2015,Tweedy2020}, durotaxis by eukaryotic cells ~\citep{roca2013mechanical,sunyer2016collective,alert2018role}, phoresis by active colloids ~\citep{Illien2017,Liebchen2018a,Stark2018}, and phototaxis by robots ~\citep{Mijalkov2016,palagireview}---exploring such effects more broadly will be a fascinating direction for future work. 

In follow-up work, the authors used a linear stability analysis of the equations governing chemotactic migration to establish the conditions under which small-amplitude morphological perturbations die away or instead grow \cite{alert2021sensing}. This analysis again reveals that the morphological stability of such chemotactic bands is determined by limitations in the
ability of individual cells to sense and thereby respond to the chemical gradient. Furthermore, it predicts that when sensing is limited at too low chemical concentrations, chemotactic bands are instead destabilized, leading to a chemotactic fingering instability that will be interesting to search for in future experiments. Intriguingly, as detailed in Ref. \citenum{alert2021sensing}, experimental data on \textit{E. coli} chemotaxis seem to suggest that the cells' sensory machinery might have evolved to ensure stable front propagation; experiments focused on
further testing this hypothesis for diverse other cell types will be useful in establishing its generality.\\

 \begin{figure}[t!]
  \begin{center}
        \includegraphics[width=\columnwidth]{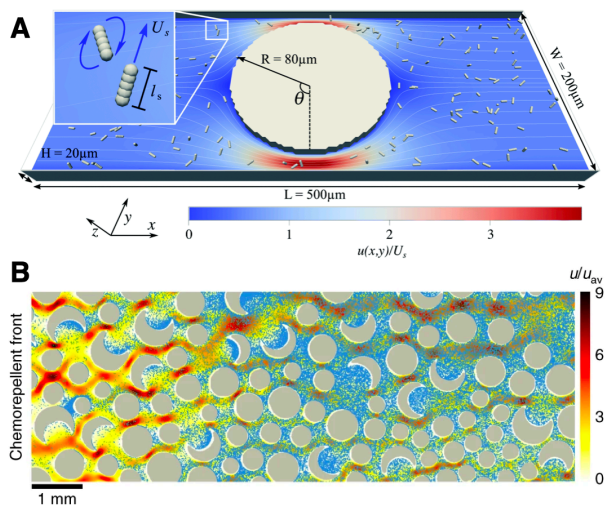}
    \caption{\begin{small}\textbf{Examples of fluid flow influencing active transport.} \textbf{A}. Lattice-Boltzmann simulation of a rectangular fluid-filled channel with a cylindrical obstacle placed at the center. Inside the channel are run-and-tumble microswimmers. The fluid is driven by an external force density. The stream lines and the colormap represent the driven flow-field projected on to the $xy$-plane normalized by the swimming speed $U_s$. Reprinted with permission from Ref.~\citenum{lee2021influence} (The Royal Society of Chemistry, 2021). \textbf{B}. Spatial distribution of \textit{Bacillus subtilis} bacteria (blue dots) in response to a chemorepellent injection, superposed on the map of fluid velocity $u$ rescaled by its average value, $u_{\text{av}}$. As a result of pore disorder, the velocity field exhibits high-velocity channels (dark red) and low-velocity micropockets (light yellow). Reprinted with permission from Ref.~\citenum{de2021chemotaxis} (Springer Nature, 2021).\end{small}\label{fig:bact-figure7}}
  \end{center}
\end{figure}

\noindent\textbf{Additional open questions.} Most studies of chemotaxis only employ a single chemoattractant, for simplicity. However, in natural habitats, bacteria are often exposed to multiple, often competing, chemical stimuli. In this case, does cellular behavior simply reflect a sum of behaviors caused by each stimulus? Building on an 1888 study by Pfeffer, Adler and Tso tested this hypothesis in 1974 by exposing \textit{E. coli} to capillary tubes containing both attractant and repellant---and concluded that the cells do indeed sum the competing behaviors ~\citep{adler1974decision}. Subsequent work, however, has suggested that this simple picture may not be complete ~\citep{strauss1995analysis}; instead, bacterial decision making in response to multiple stimuli also reflects the differing abundances of cell-surface receptors for different stimuli ~\citep{kalinin2010responses}, and thus may depend on the particular combination of stimuli presented  ~\citep{zhang2019escape}. Additionally, nutrient and chemoattractant availability can strongly fluctuate, both spatially and temporally, in many bacterial habitats ~\citep{Clausznitzer:2014:10.1371/journal.pcbi.1003870}---due to fluctuations in exogenous sources and sinks, as well as due to the secretion or degradation of chemical cues by other cells. Understanding how bacteria sense stimuli and make decisions in such complex environments, and unraveling how rich population-scale dynamics emerge from these single-cell behaviors,  will be a useful direction for future work. Furthermore, it will be interesting to develop synthetic forms of active matter that can similarly sense stimuli, process information, and respond by modulating transport in complex environments.

Many bacterial habitats also have fluid flow ~\citep{Conrad2018}, which can further alter oxygen, nutrient, and cellular profiles in interesting ways. Near surfaces, the torque from fluid shear causes the cells to rotate and swim along periodic helical-like trajectories near the surface ~\citep{rusconi2014bacterial,zottl2012nonlinear,zottl2013periodic,junot2019swimming} or even swim upstream near the surface ~\citep{figueroa2020coli,hill2007hydrodynamic,kaya2012direct,mathijssen2019oscillatory,nash2010run,tung2015emergence}, possibly promoting ~\citep{lee2021influence,secchi2020effect} or alternatively suppressing ~\citep{molaei2016succeed} attachment at specific locations. In 2D porous media, these effects can lead to retention of cells at the solid surfaces ~\citep{creppy2019effect}, but enhanced spreading of cells through the spaces between ~\citep{creppy2019effect,Alonso:2019} (Fig. \ref{fig:bact-figure7}\textbf{A}). In addition, the coupling between fluid flow and solute spreading can lead to the persistence of pore-scale chemical gradients that can lead to the retention of bacteria \textit{via} chemotaxis in low-flow regions of the pore space ~\citep{de2021chemotaxis} (Fig. \ref{fig:bact-figure7}\textbf{B}). Exploring such couplings between fluid flow, chemical transport, and bacterial transport will be an important direction for future work. Finally, we note that interactions with other stimuli, such as viscosity gradients, gravity, and light, can often play a role as well---potentially giving rise to other behaviors that motivate the development of new mathematical models ~\cite{desai2017modeling,datt2019active}.


\section{Eukaryotic cells}\label{sec:sec3}

Anyone who has observed the synchronous motion of embryonic cells or the coordinated behavior of an epithelial monolayer closing a wound can understand why the migration of eukaryotic cells has fascinated scientists for centuries. The interactions between the cells and their environment give rise to large-scale dynamics that are both fundamentally interesting, and enable the coherent migration of the group to serve a biological purpose. For example, in embryogenesis, morphogenesis, and other forms of development, cells grow, proliferate and migrate as a group in a fluid fashion through a complex environment until they reach their final shape---at which point they can collectively self-stiffen to hold their shape and withstand subsequent mechanical stresses~\citep{mongera2018fluid}, as suggested a century ago in a celebrated monograph by D'Arcy Thompson~\citep{thompson1942growth}. In addition to regulating the growth and form of living creatures, cellular migration also underlies critical pathological processes, such as tumor invasion, in which clusters of malignant cells navigate complex environments to invade healthy tissues~\citep{friedl2009collective}. 

Therefore, unraveling the role of environment-dependent physical forces and chemical signals in coupling different cells and regulating group dynamics is of fundamental importance to understanding the emergent behaviors underlying key \textit{in vivo} processes. Additionally, progress in the ability to control cell–cell interactions, tissue architecture and properties, and environmental stimuli has generated exciting possibilities to generate and control functional tissue cultures by using different guidance cues---a promising avenue that could help to develop clinically useful applications, or to engineer living meta-materials at our whim. 

In this section, we review recent studies of the transport of eukaryotic cells in environments characterized by complex geometries and external stimuli, such as electric fields, chemical gradients, and changes in mechanical properties. For other aspects and perspectives of eukaryotic cell migration, the reader is referred to the excellent reviews of Refs.~\citenum{lecuit2007cell,friedl2009collective,lecuit2011force,mayor2016front,stramer2017mechanisms,ladoux2017mechanobiology,Alert2020}.


\subsection{Complex geometries and obstacles}\label{subsec:euk-geom}


Single-cell and collective migration are strongly affected by geometric constraints. For instance, at early embryonic stages, the geometry confining the embryo is essential to build the specific cell arrangements that lead to the desired final morphology~\citep{stooke2018physical,goodwin2020mechanics,giammona2021physical}. During metastatic colonization, path-finder malignant cells outwardly migrate from the tumour colony squeezing through a microstructured extracellular matrix~\citep{kim2009tumor,ben2012bacterial,Wong2014,massague2021metastasis}. Similarly, immune cells such as macrophages, neutrophils, and dendritic cells migrate through a complex external medium, i.e. lymphoid tissues, to trap and kill pathogens~\citep{friedl2008interstitial,hampton2019lymphatic}. It is thus of great importance to understand and control how the structurally complex environments where eukaryotic cells thrive affect their main migrating mechanisms and survival strategies---a challenge that will help to understand a large number of biological and physiological processes. In this section, we highlight some of the complex structured habitats that have been explored through \textit{in vitro} experiments employing confined and intricate geometries, microchannels, and complex patterned substrata that force the cells to adopt a certain shape.\\

\begin{figure*}[t!]
  \begin{center}
        \includegraphics[width=0.7\textwidth]{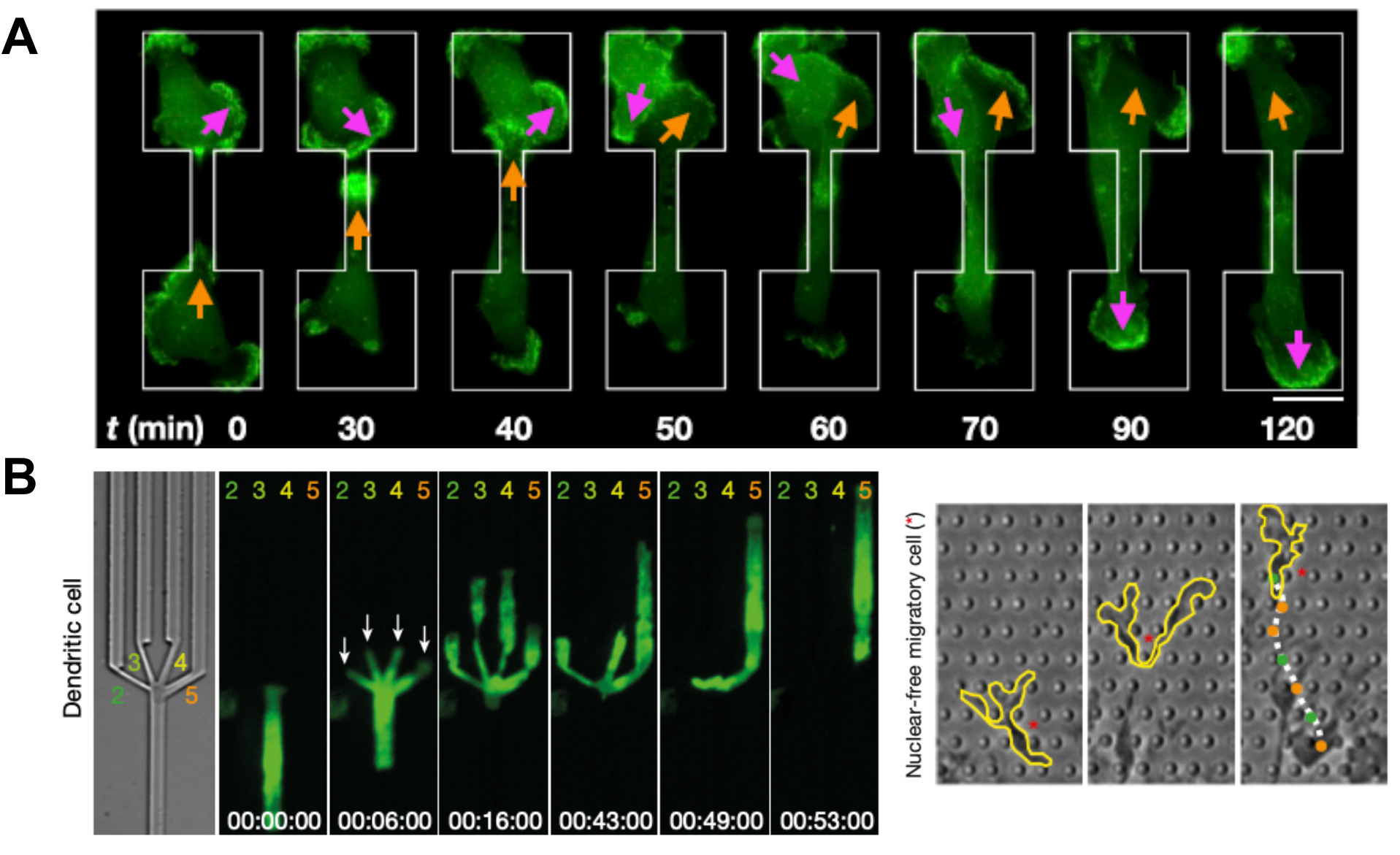}
    \caption{\begin{small}\textbf{Examples of single-cell migration in structured 2D environments}. \textbf{A}. Time-lapse snapshots showing the transient dynamics of two eukaryotic cells (MDA-MB-231 cells) interacting inside a confined geometry. Arrows indicate regions of high actin activity. Scale bar is $25~\mu$m. Reprinted with permission from Ref.~\citenum{bruckner2021learning} (National Academy of Sciences, 2021). \textbf{B}. Left: Dendritic cell choosing the path of least resistance within an confined geometry with pores of different sizes. Right: nucleus-free cytoplasts (outlined in yellow) migrating through a pillar array of large (orange) and small (green) pores without choosing any preferred path. Time is indicated as h:min:s. Adapted with permission from Ref.~\citenum{renkawitz2019nuclear} (Springer Nature, 2019).\label{fig:figure1}\end{small}}
  \end{center}
\end{figure*}

\noindent \textbf{Single-cell migration in structured 2D environments.} The motion of a single cell is a well-studied process that is dictated by intra/extracellular signals and mechanical forces~\citep{ridley2003cell}---yet many aspects involving cell-cell interactions and the role of a complex confining environment on their dynamics remain poorly understood. Concerning 2D configurations, several works have explored the effect of structured and intricate geometries, together with solid obstacles on the behavior and migration of single eukaryotic cells. Fig.~\ref{fig:figure1} highlights some of these studies.

One simple way of imposing geometric constraints is using ``two-state" micropatterns consisting of two square adhesive sites connected by a thin gap. Ref.~\citenum{bruckner2019stochastic} studied the single-cell migration dynamics of cancerous (MDA-MB-231) and a non-cancerous (MCF10A) cells in such geometries. The authors found that the motions of both type of cells exhibit qualitatively similar nonlinear migratory dynamics \com{in the position-velocity cellular phase space resembling a nonlinear oscillator}, which can indeed be decomposed into deterministic and stochastic contributions. In particular, they showed that these two cells exhibit different deterministic dynamics---namely, cancerous cells reached a limit cycle, \com{i.e. the position and velocity amplitudes of the self-sustained oscillations reach a fixed value at long time,} and non-cancerous cells displayed excitable bistable dynamics. \com{In this latter scenario, the migratory dynamics does not exhibit self-sustained oscillations, and therefore instead of a limit cycle the authors found that two fixed points for the position and velocity appeared on either side of the bridge. They found that the basins of attraction of these two fixed points occupied all the micro-environment, thus any small perturbation can trigger a switch from one state to the other, indicating the excitable bistable dynamics of MCF10A cells}. More recently, Ref.~\citenum{bruckner2021learning} studied cell-cell interactions and cell coordination using the same confined micro-environment (Fig.~\ref{fig:figure1}\textbf{A}), which the authors referred to as an \textit{experimental cell collider}. They observed \com{several} collision behaviors depending on the interaction between different types of cells. In particular, repulsion and friction interactions dominate between non-cancerous cells, whereas attractive and anti-friction interactions arise between cancerous cells. In the former scenario, cells slow down as they move past each other, which corresponds to a reversal behavior usually termed as `contact inhibition' of locomotion. In the latter, the so-called antifriction produces a sliding behavior that the authors termed `contact sliding' locomotion, where cancerous cells accelerate as they move side by side with increasing velocity, becoming weakly repulsive at long distances. Both distinguished dynamics are in fairly good agreement with their theoretical results based on an interacting stochastic equation of motion. With this theory, the authors also found another interaction dynamics where cells follow each other, a behavior usually referred to as `contact following' locomotion. This work exemplifies how analyzing cell-cell interactions in engineered micro-environments can help shed light on the mechanisms by which cancerous and non-cancerous cells interact during \textit{in vivo} processes---such as the migratory dynamics of malignant cells during tumor invasion, in which the interactions between cancerous and non-cancerous cells underlie the competition between both populations.

 The migration of macrophages, neutrophils, and dendritic cells through complex lymphoid tissues plays a crucial role during immune surveillance and immune responses. To unravel the behavior of immune cells in such complex environments, Ref.~\citenum{renkawitz2019nuclear} studied their migration dynamics in different 2D and 3D geometries. Concerning the former, the authors analyzed the migration of individual dendritic cells moving within intricate branching microchannels consisting of a main channel splitting into four different sized smaller channels (see Fig.~\ref{fig:figure1}\textbf{B}). The authors observed that the cells migrate through the main channel until the branching point. Strikingly, the cells then spread their bodies through the different branches, using their nucleus as an indicator to choose the best path, thus being able to move through the path of least resistance (Fig.~\ref{fig:figure1}\textbf{B}, left);  nucleus-free cytoplasts, on the other hand,
 do not select any particular path when navigating through arrays of cylindrical pillars (Fig.~\ref{fig:figure1}\textbf{B}, right). By highlighting the role of the nucleus as a sensor of the cellular microenvironment, this research thus helps to shed light on the process by which immune cells help navigate through the body and disseminate malignant tumor cells.\\

\begin{figure*}[htp]
  \begin{center}
        \includegraphics[width=0.70\textwidth]{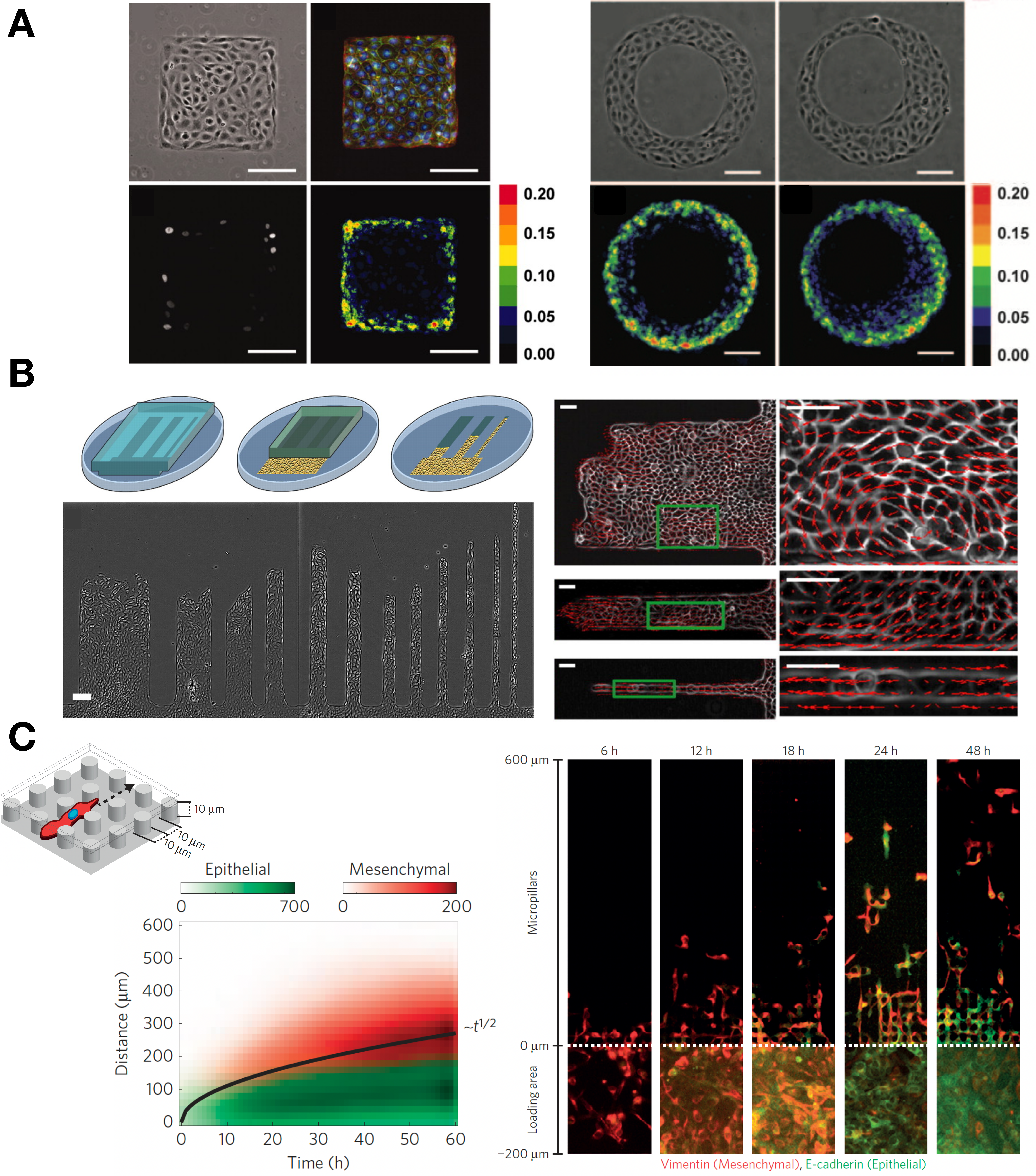}
    \caption{\begin{small}\textbf{Collective cell migration in structured 2D environments}. \textbf{A}. Substrate patterning shows that eukaryotic cells (bovine pulmonary artery endothelial cells) experience different proliferation patterns at the edges than in the bulk. Scale bars are $10~\mu$m. Reprinted with permission from Ref.~\citenum{nelson2005emergent} (Copyright National Academy of Sciences, 2005). \textbf{B}. Left: sketch of the experimental setup, and two snapshots displaying the collective migration of MDCK cells in micro-printed stripes. Scale bar is $100~\mu$m. Right: bulk instabilities, namely swirls and vortices, disappear under strong confinement, i.e. in the narrow stripes. Scale bars are $50~\mu$m. Reprinted with permission from Ref.~\citenum{vedula2012emerging} (National Academy of Sciences, 2012). \textbf{C}. Left: sketch of the experimental setup consisting of an array of micropillars where uniform populations of mammary epithelial cells (MCF-10A) and breast adenocarcinoma cells (MDA-MB-231) were cultured. Right and bottom: EMT-activated cells migrate collectively advancing as a front where individual cells detached and move individually. Below the front, cells exhibited phenotypic plasticity undergoing a MET transition. Color bars indicate the number of epithelial and mesenchymal cells as functions of time and the distance from the initial migration front. Adapted with permission from Ref.~\citenum{Wong2014} (Springer Nature, 2014). \end{small}\label{fig:figure3}}
  \end{center}
\end{figure*}

\noindent \textbf{Collective migration in structured 2D environments.} In most \textit{in vivo} situations cells grow and migrate as a group. Hence, a growing area of research focuses on the collective behavior of groups of cells in complex 2D environments mimicking those arising in fundamental biological processes, such as morphogenesis, development, wound healing, and cancer progression. In Fig.~\ref{fig:figure3}, we highlight different works that have explored the collective behavior of eukaryotic cells in structured 2D geometries.

One way of creating complex 2D environments is using substratum patterning using adhesive islands coated with extracellular matrix~\citep{chen1997geometric}. A collective of eukaryotic cells can then grow and adapt their overall shape to that of the adhesive islands. For isolated cells, the striking results of Ref.~\citenum{chen1997geometric} showed that cell shape directly determines life and death, and proliferation and quiescence, although the exact mechanisms by which cells transduce changes in their geometry to biochemical responses are still the subject of active inquiry. Then, using aggregates of bovine pulmonary artery endothelial cells and normal rat kidney epithelial cells on microprinted substrata of different shapes and sizes, the authors of Ref.~\citenum{nelson2005emergent} showed that the geometry of the substratum pattern tunes the mechanical forces associated with a growing monolayer, which in turn determines the resulting patterns of cellular proliferation (\com{Fig.~\ref{fig:figure3}\textbf{A}}). Further increasing the complexity of the 2D environment, the authors of Ref.~\citenum{vedula2012emerging} then cultured 2D monolayers of Madin-Darby Canine Kidney (MDCK) epithelial cells in micro-printed stripes, as shown in \com{Fig.~\ref{fig:figure3}\textbf{B}}. The authors found that, under strong confinement in narrow stripes, collective motions such as multicellular swirls and vortices that typically arise in monolayers in uniform, unstructured environments disappear. Instead, the monolayer only migrates unidirectionally \textit{via} a contraction-relaxation mechanism.

\begin{figure*}[t!]
  \begin{center}
        \includegraphics[width=0.7\textwidth]{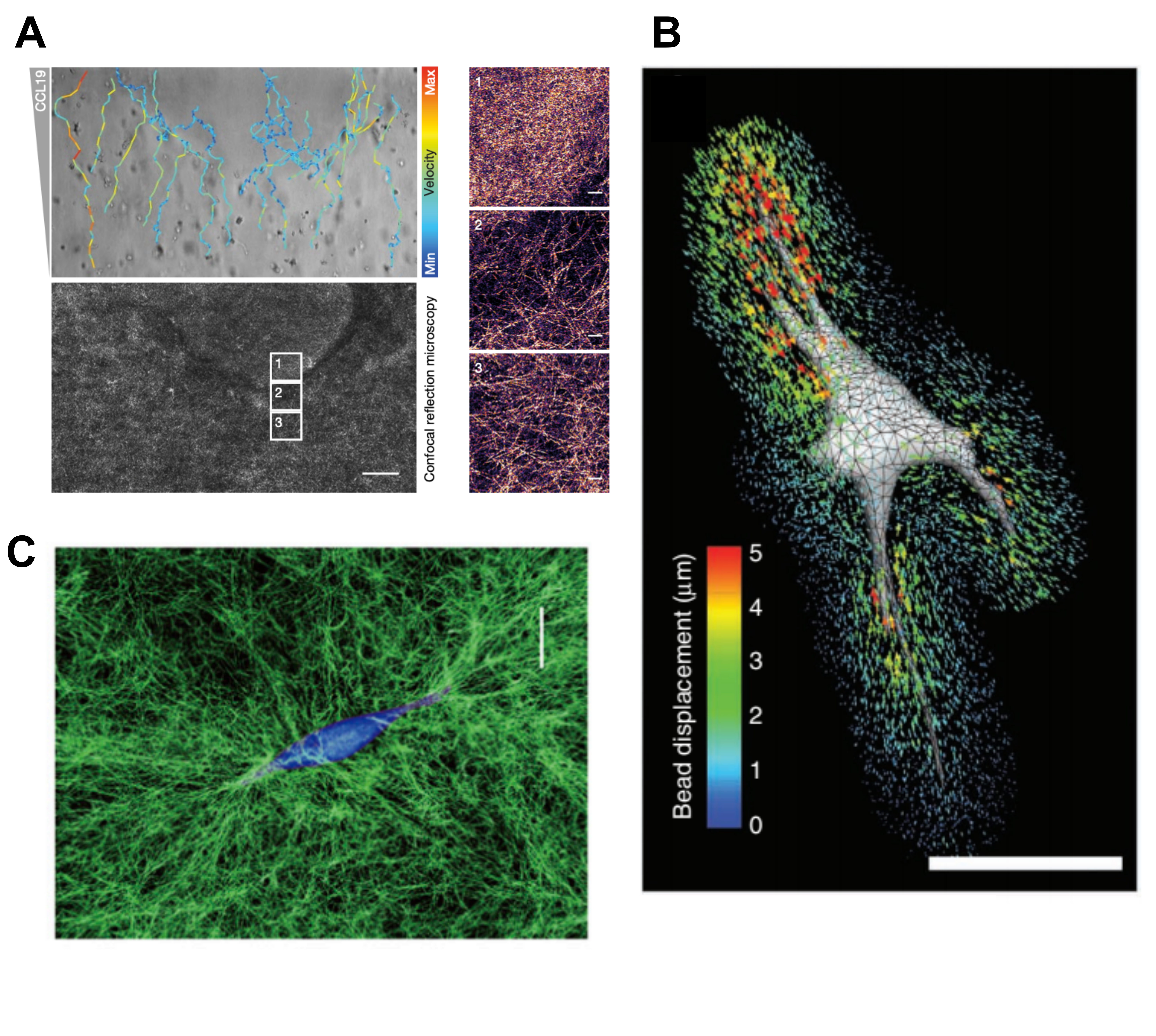}
    \caption{\begin{small}\textbf{Examples of individual eukaryotic cells interacting with complex 3D environments}. \textbf{A}. Dendritic cells migrating through a complex 3D collagen matrix. Cells move towards the region of higher collagen density \textit{via} chemotaxis. The cells that encountered high density regions tangentially circumvented them, thus choosing the path of least resistance. Reprinted with permission from Ref.~\citenum{renkawitz2019nuclear} (Springer Nature, 2019). Scale bar on the left is $100~\mu$m and on the enlarged regions on the right $10~\mu$m.
    \textbf{B}. Deformation of a hydrogel induced by an encapsulated fibroblast. In the color bar the bead displacement trajectories are represented. Scale bar is $50~\mu$m. Reprinted with permission from Ref.~\citenum{legant2010measurement} (Springer Nature, 2010). \textbf{C}. An individual MDA-MB-231 cell in blue contracting a 3D collagen network in green. Scale bar is $10~\mu$m. Reprinted with permission from Ref.~\citenum{han2018cell} (National Academy of Sciences, 2018).\end{small} \label{fig:figure2}}
  \end{center}
\end{figure*}

Individual and collective cell migration also plays a crucial role during pathological processes such as tumor invasion and metastasis. In cancer progression, it is known that malignant cells undergo different phenotypic transitions and acquire characteristics of embryonic mesenchymal cells, thus becoming loosely packed and less organized, and exhibiting a greater migratory capacity that enables invasion. This phenomenon is usually referred to as the epithelial-mesenchymal transition (EMT)~\citep{yang2004twist,polyak2009transitions,thiery2009epithelial,kalluri2009basics,hanahan2011hallmarks,francou2020epithelial}. When these malignant cells detach from the tumor community, they adapt their migratory path to the complex medium they inhabit and invade healthy tissues and organs far from the primary tumor. To understand this change in behaviors, the authors of Ref.~\citenum{Wong2014} studied the migration of \com{uniform cultures of} epithelial (\com{mammary epithelial MCF-10A}) and mesenchymal (\com{breast adenocarcinoma MDA-MB-23}) cells in a micro-fabricated 2D array of pillars \com{(Fig.~\ref{fig:figure3}\textbf{C})}. Intriguingly, they found that uniform cultures of breast adenocarcinoma MDA-MB-231 mesenchymal cells eventually \com{exhibit a phenotypic plasticity in the loading area with no pillars (region $y < 0$ in Fig.~\ref{fig:figure3}\textbf{C}), thus} acquiring epithelial \com{characteristics} associated with a mesenchymal-epithelial transition (MET); however, the moving interfacial front scattered EMT-activated cells that migrate individually through the pillars with a larger velocity than the collective migration. \com{The authors showed that the migration front propagates as the square root of time, and proposed an interesting analogy comparing such emergent dynamics with a model inspired by phase transitions during binary-mixture solidification. In their experiments, the solidification advancing front is associated with the MET in the loading region. The authors clearly showed that the scattering behavior at the collectively advancing front is enhanced by the regularly spaced micro-pillars, since they disrupt cell-cell contact stimulating individual cell migration, both characteristics associated with EMT. This work evidences that analyzing the collective dynamics of cells in two-dimensional structured environments can provide quantitative measurements and new insights into tumor dissemination during metastasis. Additionlly, this research direction could pave the way to designing new therapeutic strategies that inhibit individual migration during tumor invasion by enhancing MET.}\\


\noindent\textbf{Single-cell migration in structured 3D environments.} The healthy and pathological processes described above, namely immune surveillance, wound healing, or tumor invasion, typically occur within a heterogeneous 3D environment crowded by cells and extracellular matrix. Although several studies have unraveled the molecular and biophysical mechanisms involved in single and collective cell migration in 2D configurations, as briefly highlighted above, how cells navigate through complex 3D tissue structures remains poorly understood~\citep{wu2018biophysics,yamada2019mechanisms,Alert2020}. This gap in knowledge arises partly due to the new modes of motility that may emerge in structured 3D environments, since they could be a combination of the different migration modes employed by eukaryotic cells, e.g., amoeboid (fast migration mode in which a rounded cell with a leading-edge protrusion undergoes deformation to squeeze through gaps), mesenchymal (slower migration mode in which a more polarized elongated cell develops actin-rich leading-edge protrusions that produce adhesive interactions with the medium), or collective (cells migrating coherently as a group), depending on the context~\citep{friedl2009collective,wirtz2011physics,roussos2011chemotaxis,paul2016engineered,yamada2019mechanisms}. Additionally, the interaction with a compliant 3D environment is considerably different compared from the case of a rigid 2D substrate. For instance, in 3D, focal adhesions to collagen fibers are smaller and are known to regulate the cellular migration speed; moreover, the subcellular cytoskeleton appears to be organized differently, and thus cell-exerted tractions are also different~\citep{bloom2008mapping,fraley2010distinctive,legant2010measurement,han2018cell,yamada2019mechanisms}. Exploring first the migration of single eukaryotic cells within complex 3D environments is thus crucial for quantitative insights into the biological processes cited above.


In the context of immune cells, the authors of Ref.~\citenum{renkawitz2019nuclear} explored the migration of dendritic cells through heterogenous hydrogels with pore sizes much smaller than the size of a single cell. In particular, the authors investigated the migration of immune cells in collagen networks with graded pore sizes ranging from 1 $\mu$m to 5 $\mu$m arising from gradients in the density of collagen fibers. To this end, they forced the cells to migrate in response to chemotactic gradients (detailed further in the next section) directed towards regions of smaller pores size/higher collagen density. As shown in \com{Fig.~\ref{fig:figure2}\textbf{A}}, cells facing head-on regions of smaller pore size/higher density continue moving aligned with the chemotactic cue, whereas cells encountering those regions more tangentially circumnavigated the smaller pore size/high-density region by deviating from the chemotactic path. Together with the findings in 2D confined geometries described previously, these results suggest that cells have the ability to distinguish between different pore sizes, migrating preferentially towards the path of least resistance.

Additionally, several works have performed quantitative force measurements of isolated cells embedded in 3D matrices to discover how eukaryotic cells mechanically interact with their environment~\citep{bloom2008mapping,fraley2010distinctive,legant2010measurement,han2018cell}. In Ref.~\citenum{legant2010measurement} the authors described a technique that measures the traction stresses exerted by cells in 3D elastic hydrogels by tracking the displacement of fluorescent beads in the vicinity of each cell (\com{Fig.~\ref{fig:figure2}\textbf{B}}). Using this methodology, the authors showed that isolated cells probe their surrounding environment by exerting strong inward tractions using long and slender extensions. More recently, building on this work, the authors of Ref.~\citenum{han2018cell} used a combination of theory and numerical simulations to demonstrate that a related measurement technique based on nonlinear stress inference microscopy captures the cell-induced stress field. Experiments on isolated cells embedded in different 3D extracellular matrices---collagen, fibrin,
and Matrigel---revealed that cells interact with their surrounding network by pulling on it, producing large stresses that in turn generate large stiffness gradients (\com{Fig.~\ref{fig:figure2}\textbf{C}}). \\

\begin{figure*}[htp]
  \begin{center}
        \includegraphics[width=0.7\textwidth]{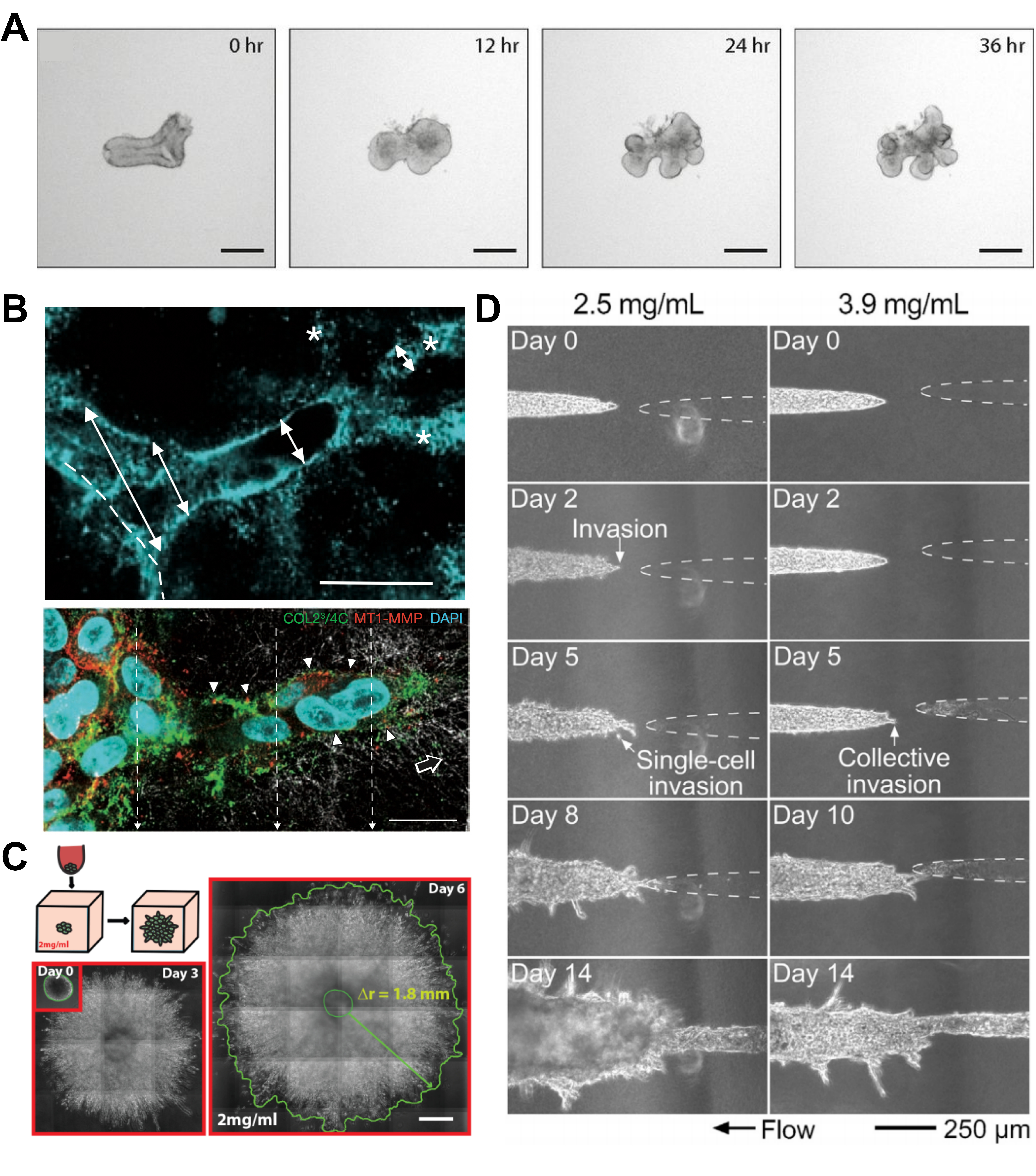}
    \caption{\begin{small}\textbf{Examples of collective cell migration and growth in structured 3D environments}. \textbf{A}. Time-lapse images showing the growth and branching of mesenchyme-free mouse lung embryonic epithelium embedded in a 3D Matrigel. Reprinted with permission from Ref.~\citenum{varner2015mechanically} (National Academy of Sciences, 2015). Scale bar is $200~\mu$m. 
     \textbf{B}. \textit{In vitro} multicellular invasion through a 3D collagen matrix by HT-MT1 cells (fibrosarcoma). Cells migrate collectively by degrading the collagen matrix. The arrow indicates the migration direction, and scale bar is $20~\mu$m. Reprinted with permission from Ref.~\citenum{wolf2007multi} (Springer Nature, 2007). \textbf{C}. Experimental setup and time-lapse images of a tumor spheroid (HT1080 human fibroscarcoma) invading a type I collagen matrix at 0, 3 and 6 days. Scale bar is $500~\mu$m. Reprinted with permission from  Ref.~\citenum{valencia2015collective} (Impact Journals, 2015).
    \textbf{D}. 3D microtumor embedded in collagen gels of different concentration. At low-density collagen gels (left column), MDA-MB-231 breast cancer cells invade the cavity faster than in high-density collagen matrices (right column). At low-density gels the invasion started even before interstitial flow was applied and single-cell and collective migration coexisted. Reprinted with permission from Ref.~\citenum{tien2020matrix}   (Elsevier, 2020).\end{small}
\label{fig:figure4}}
  \end{center}
\end{figure*}


\noindent\textbf{Collective migration in structured 3D environments.} 
Molecular signals and resulting biochemical patterns are typically thought to dictate the spatial patterns that emerge during embryonic development and morphogenesis. Nevertheless, several works have shown that physical forces in structured 3D environments can also drive spatial patterning of cellular groups during developmental processes, as detailed further in Refs.~\citenum{stooke2018physical,goodwin2020mechanics}. For example, the authors of Ref.~\citenum{varner2015mechanically} showed that even in the absence of mesenchyme, \com{a loosely organized embryonic connective tissue of undifferentiated cells that surrounds the epithelium and helps the formation of spatial patterns}, purely physical mechanisms are able to determine the spatial patterning of an embryonic airway epithelium \com{(see Fig.~\ref{fig:figure4}\textbf{A})}. In particular, the authors analyzed the branching morphogenesis of a mesenchyme-free embryonic lung embedded in a viscoelastic 3D gel matrix (Matrigel). They showed that the dominant wavelength of the branching patterns varies with the gel concentration, observing that explants cultured within higher concentrations of Matrigel displayed shorter branch wavelengths. Since the rate of proliferation also increased with the gel concentration, due to the growth factors present at Matrigel, the authors disentangled the role of the gel mechanical properties and the epithelial growth on the branching patterns. In particular, they showed that the experimental results were independent of the gel stiffness, i.e. altering the matrix stiffness with methylcellulose while keeping constant the ligand density did not affect the branch wavelengths, thus suggesting that a purely elastic instability is not responsible for the pattern formation. However, increasing the fibroblast growth factors while keeping the mechanical properties of Matrigel constant, increased the epithelial expansion and decreased the dominant wavelength, suggesting that a growth-induced viscoelastic instability is the responsible for the spatial patterning, in agreement with a simplified viscoelastic Maxwell model showing that the dominant wavelength is independent of the mechanical properties of the surroundings.

Concerning tumor invasion and dissemination of malignant cells, several works have analyzed how a complex 3D environment, such as a collagen matrix, affects the motility and invasion of malignant cells. For example, Ref.~\citenum{wolf2007multi} studied the collective 3D migration of fibrosarcoma (HT-MT1) and breast cancer (MDA-MB-231) cells through a 3D collagen matrix, as shown in \com{Fig.~\ref{fig:figure4}\textbf{B}}. Notably, invasive migration through these complex 3D media exhibits significantly different features to that of migration on 2D substrata. In 3D matrices, collective cell migration requires pericellular remodelling of the extracellular matrix, in which cells \com{located at the edge of the colony} locally degrade the matrix by secreting metalloproteases to eventually migrate collectively in cellular strands. Similarly, the authors of Ref.~\citenum{valencia2015collective} performed experiments on migrating HT1080 human fibroscarcoma spheroids invading a 3D collagen matrix (\com{Fig.~\ref{fig:figure4}\textbf{C})}. They showed that tumor cells at the periphery \com{of the spheroidal tumor colony} become polarized, adopting an aligned and elongated morphology, which enables them to migrate more effectively than cells located closer to the core where cell density is higer. Their results suggest that this directed cell dissemination at the \com{border of the tumor community} is a general feature of migration in 3D collagen matrices, which requires integrin-based cell-collagen adhesion and myosin-based contractility. Taken together, these works show that finger-like instabilities at the edge of the invasive front not only arise in 2D~\cite{omelchenko2003,poujade2007,sepulveda2013,basan2013,tarle2015modeling,mayor2016front,begnaud2016mechanics,nesbitt2017edge,vishwakarma2018mechanical,williamson2018stability,Alert2019,xi2019material,yang2020leader,bonilla2020tracking,trenado2021cellsubstrate}, but also in 3D environments. More recently, by using a 3D microfluidic culture system, the authors of Ref.~\citenum{tien2020matrix} showed that low-density collagen gels hasten the invasion of MDA-MB-231 breast cancer cells---a phenomenon that only depends on the pore size, thus being independent of the interstitial flow speed and the elastic properties of the surrounding medium. Furthermore, in gels of low collagen density, cellular invasion towards a lymphatic-like cavity started even in the absence of an applied interstitial flow, with both single and collective cell migration operating simultaneously (\com{see Fig.~\ref{fig:figure4}\textbf{D})}.

Thus, interactions with a structured 3D environment regulate both single-cell and collective migration and growth. Extending existing theoretical frameworks, and developing new ones, to treat these interactions will be an important direction for future work. For example, although challenging, discrete models, such as agent-based or active-vertex models accounting for sub- and supra-cellular features~\citep{marchetti2013hydrodynamics,hakim2017collective,barton2017active,Alert2020}, can be extended to account for a structured environment, namely intricate and/or compliant geometries. Additionally, despite the complexity of these systems, continuum theories based on liquid crystal~\citep{de1993physics} and active-gel physics~\citep{marchetti2013hydrodynamics,prost2015active} not only have advanced our understanding of single-cell behavior, but have been also employed to describe the physics of collective eukaryotic cells by coarse graining different supra-cellular mechanisms~\citep{marchetti2013hydrodynamics,hakim2017collective,Alert2020}. Incorporating features of complex and structured environments into such frameworks could provide a way to predict cellular behavior more generally.


\begin{figure*}[t!]
  \begin{center}
        \includegraphics[width=0.7\textwidth]{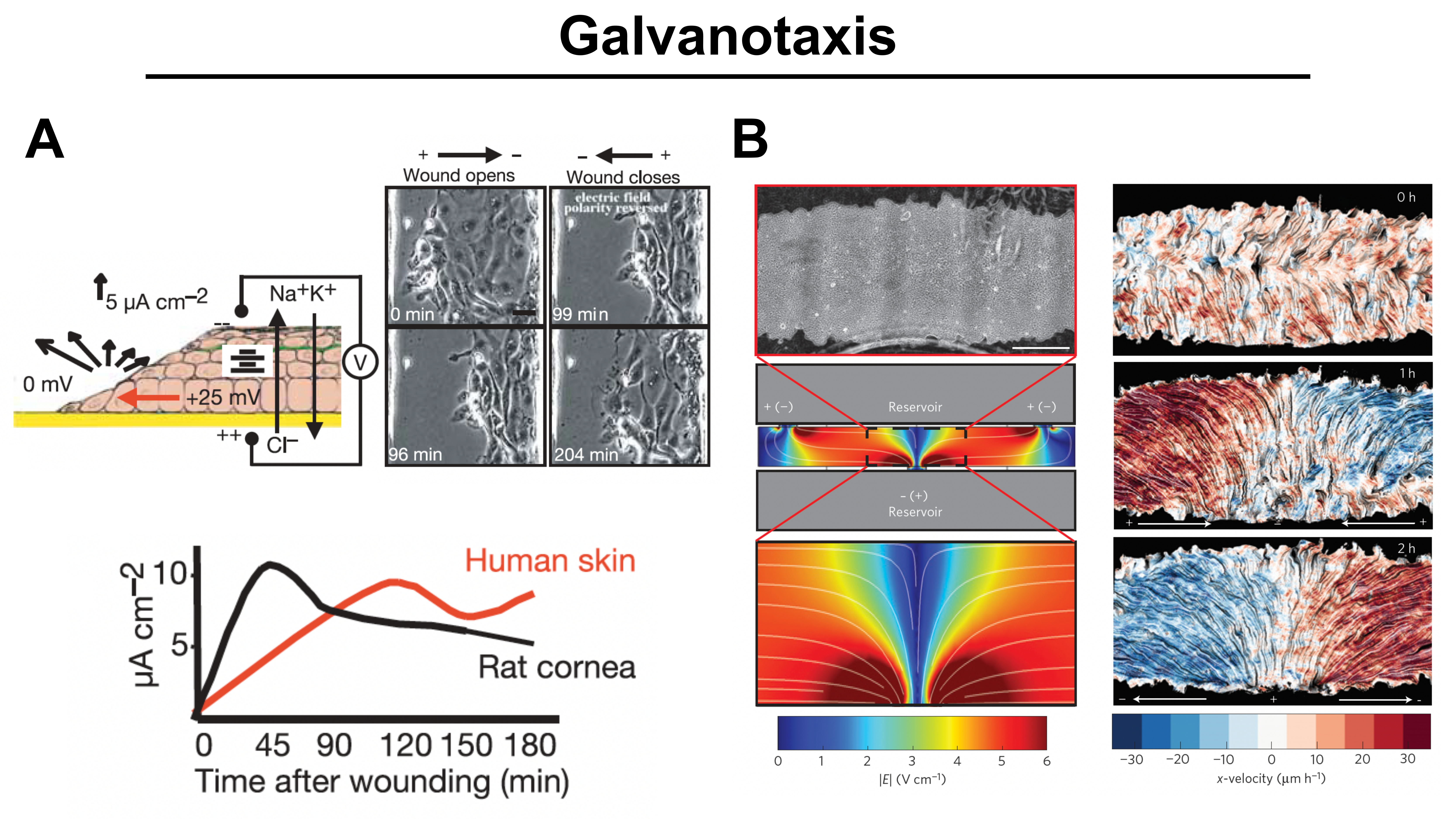}
    \caption{\begin{small}\textbf{Examples of galvanotaxis influencing the collective migration of eukaryotic cells}. \textbf{A}. Left: sketch of the monolayer model of wound healing and the electric current as a function of time during healing for human skin and rat cornea cells. The red arrow indicates the endogenous lateral electric field directed towards the wound due to the trans-epithelial potential difference. Right: electric-field induced cell migration. The wound opens or closes depending on the polarity of the electric field. Scale bar is $20~\mu$m. Reprinted with permission from Ref.~\citenum{zhao2006electrical} (Springer Nature, 2006). \textbf{B}. Left: top view of the experimental chamber and a color map indicating the magnitude of the divergent electric field. Right: Epithelial monolayer behavior in terms of the horizontal ($x$) velocity after the application of an electric field. Scale bar is $800~\mu$m. Reprinted with permission from Ref.~\citenum{cohen2014galvanotactic} (Springer Nature, 2014).\end{small}\label{fig:figure5}}
  \end{center}
\end{figure*}

\subsection{External stimuli}
A strong scientific effort has been devoted to unravel the emergent behaviors of eukaryotic cells when they are exposed to diverse external stimuli. In this section, we review some of the most studied external stimuli and guidance cues applied to eukaryotic cells---namely electric fields (galvanotaxis), chemical gradients (chemotaxis), or gradients in the substrate stiffness (durotaxis). Other types of taxis and guidance cues are described in e.g., Refs.~\citenum{roca2013mechanical,cortese2014influence,haeger2015collective}.\\




\noindent\textbf{Galvanotaxis.} It is well known that, under normal physiological conditions, eukaryotic cells generate a voltage across their plasma membrane, usually referred to as the membrane potential. This voltage is used for several intracellular functions including transport and signaling. Moreover, at the collective cell level, tissues, organs, and embryos are surrounded by an epithelial layer responsible for producing trans-epithelial potentials. Hence, electric fields are endogenous and carry out fundamental functions during several biological processes such as tissue repair, wound healing, and embryonic development~\citep{jaffe1977electrical,nuccitelli2003role,mccaig2005controlling,zhao2006electrical,cortese2014influence,ross2017use}.

Furthermore, for over a century, it has been known that eukaryotic cells exposed to an external d.c. electric field similar in magnitude to those detected endogenously respond by reorienting and migrating in the direction dictated by the electric potential---a process known as \textit{galvanotaxis}. Indeed, this research direction dates back to the pioneering works of L. Galvani, C. Mateucci, and E. duBois-Reymond~\citep{du1843vorlaufiger}, who measured the electric field in injuries and during wound healing, during the 18th and 19th centuries. In particular, they found that when a wound occurs, the disruption in multicellular integrity triggers an electric field of $~$0.1-1 V cm$^{-1}$ to be directed towards the wound. In his seminal work in 1891~\citep{dineur1891note}, E. Dineur subsequently found that leukocytes exhibit a directed motion guided by the electric current, a phenomenon that he coined as galvanotaxis in honor of L. Galvani. Despite these early discoveries, the precise mechanisms underlying this form of taxis are still under debate~\citep{allen2013electrophoresis,cortese2014influence,ross2017use} in large part due to its dependence on the cell type, medium conditions, and on different signaling cascades. However, researchers are converging on the idea that electrotactic and chemotatic cues share common downstream motility pathways~\citep{zhao2006electrical,allen2013electrophoresis,sun2013keratocyte}. Thus, it is now well established that electric cues have great potential in directing collective cell migration and controlling emergent group behaviors~\citep{zhao2006electrical,mccaig2005controlling,cortese2014influence,cohen2014galvanotactic,ross2017use}, which could be used eventually for beneficial purposes in clinical applications. Indeed, galvanotaxis provides more precise control and faster responses than other external stimuli, such as chemical signals. 





Concerning spreading epithelial monolayers and wound healing, the experiments of Ref.~\citenum{zhao2006electrical} showed that electric fields are able to override other guidance cues. The authors showed that an external d.c. electric field is able to alter wound healing, enhancing healing when is applied in the same direction of the trans-epithelial potential, or opening the wound when is directed in the opposite direction (Fig.~\ref{fig:figure5}\textbf{A}). Their results also suggest that electrotactic and chemotatic cues activate common signaling pathways. More recently, the work of Ref.~\citenum{cohen2014galvanotactic} studied the collective cell migration of a 2D epithelial monolayer undergoing galvanotaxis (Fig.~\ref{fig:figure5} \textbf{B}). The authors found that, after applying a divergent electric field for one hour, the migration of the whole epithelial monolayer was guided by the electric current density. Strikingly, the group of cells moved collectively in the direction aligned with the electric-field lines while maintaining the epithelial integrity, as show in Fig.~\ref{fig:figure5}\textbf{B}. By contrast, due to the characteristic behavior of the leader cells located at the edge of the monolayer, the electric field was not able to polarize them and guide their migration. Building on these results, the authors of ~\citep{zajdel2020scheepdog,zajdel2020come,shim2021cellular} developed an experimental device to analyze the behavior of large tissues undergoing galvanotaxis. Using this device, they found that wound healing is accelerated when an external electric field is applied, the healing process being twice faster than under normal conditions. Additionally, strongly-adhesive tissues are more complex to control \textit{via} galvanotaxis; nevertheless, a sufficiently strong electric field can eventually damage the epithelium. 

\begin{figure*}
  \begin{center}
        \includegraphics[width=0.7\textwidth]{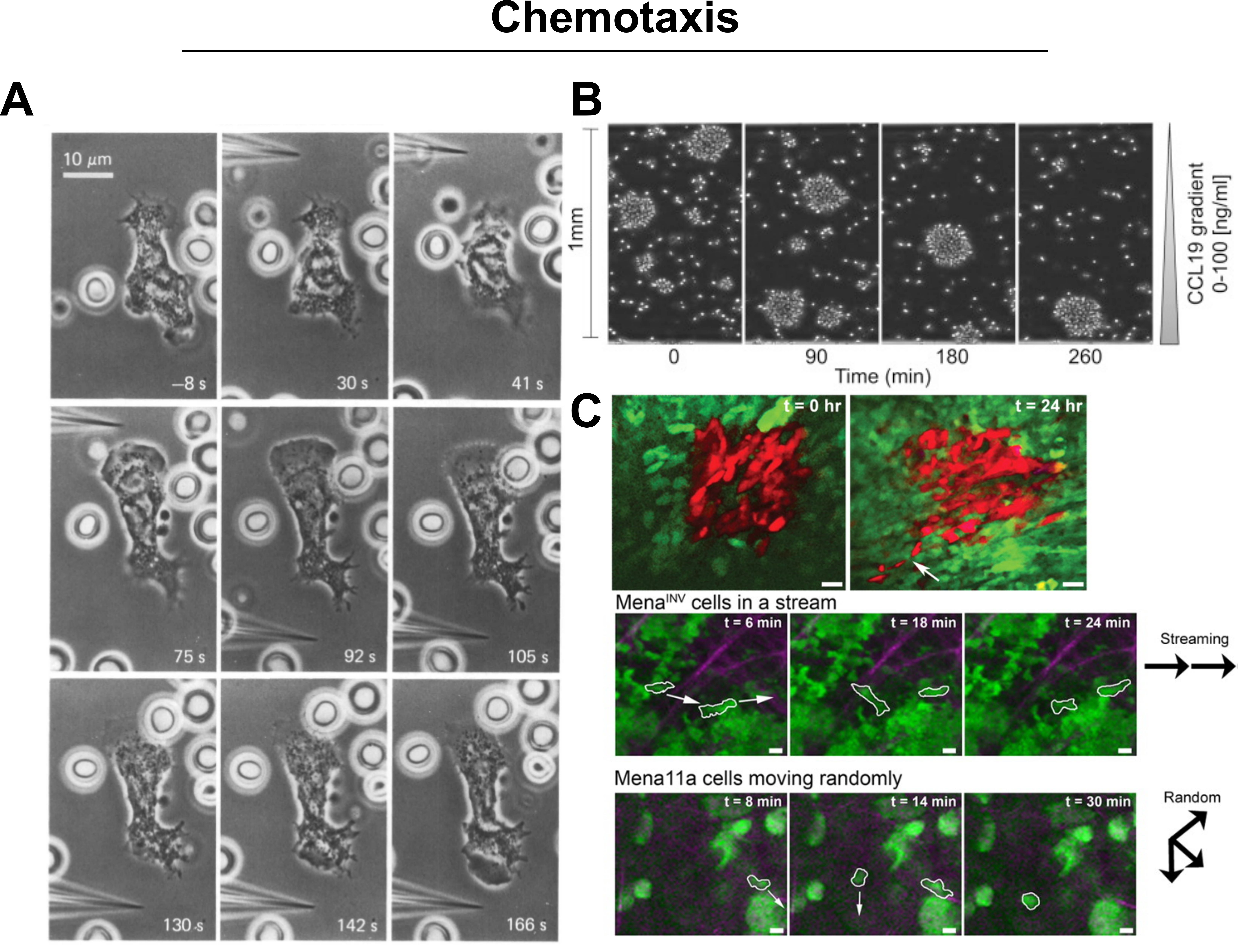}
    \caption{\begin{small}\textbf{Examples of chemotactic guidance in both individual and collective cell migration within structured environments}. \textbf{A}. Chemotactic reorientation of a granulocyte towards a stimulus (pipette). Reprinted with permission from Ref.~\citenum{gerisch1981chemotactic} (The Company of Biologists, 1981). \textbf{B}. Single cells and cluster of cells migrating toward the chemoattractant source in a very directional manner. Reprinted with permission from Ref.~\citenum{malet2015collective} (Elsevier, 2015). \textbf{C}. Upper panel: Long-lived behavior of coordinated movement in carcinoma cell streams. Scale bars are $50~\mu$m. Middle panel: coordinated motion of cells through the tumor position (indicated with the arrows). Scale bars are $25~\mu$m. Lower panel: Random motion of cells in presence of a tumor. Scale bars are $25~\mu$m. Reprinted with permission from Ref.~\citenum{roussos2011mena} (The Company of Biologists, 2011).\end{small}\label{fig:figure6}}
  \end{center}
\end{figure*}
From a theoretical perspective, the coupling between an external d.c. electric field and the stresses involved during eukaryotic collective migration, such as active traction stresses, contractile stresses, and viscous/elastic stresses~\citep{Alert2020}, remains underexplored. A systematic comparison between experiments and new discrete and continuum models incorporating electric stresses and current-dependent polarization will be essential to unravel the physics involved in galvanotaxis and the collective behaviors that result.\\







\noindent\textbf{Chemotaxis.} Eukaryotic cells also have the ability to sense and respond to extracellular chemical stimuli, although the underlying mechanisms employed are  distinct from  chemotaxis in prokaryotic organisms~\citep{ridley2003cell}, discussed previously in Section~\ref{bac-stimuli}. Within the context of eukaryotic organisms, chemotaxis is the most extensively studied directional cue, mainly due to its fundamental role in a large variety of biological processes. For instance, chemotaxis plays a key role in driving individual and collective cell migration in developmental processes. It is also important for the trafficking of immune cells and in tissue regeneration/wound healing. Moreover, chemotaxis is not only relevant in healthy physiological processes, but it also plays key roles in the progression cancer~\citep{muller2001involvement,kedrin2007cell,roussos2011chemotaxis}, as well as in the pathogenesis of inflammatory diseases such as asthma, allergies, and atherosclerosis~\citep{luster1998chemokines}.

The first reported observations of chemotaxis in eukaryotic cells date back to the 19th century by the ophthalmologist T. Leber~\citep{leber}, who showed the attraction of leukocytes by chemical gradients, although the precise extracellular mechanism responsible for guiding leukocytes was only identified several decades later~\citep{zigmond1977ability,devreotes1988chemotaxis,jin2008chemotaxis}. In 1947, J.T. Bonner and L.J. Savage found that the soil amoeba \textit{Dictyostelium discoideum} responds to the gradient of a chemical by deforming and migrating in the direction of the gradient, thus clearly showing that chemical signals are able to polarize and organize the cellular movement~\citep{bonner1947evidence}. This work paved the way for subsequent work utilizing \textit{D. discoideum} as a model organism to study chemotaxis~\citep{gerisch1968cell,gerisch1981chemotactic,gerisch1981chemotactic,devreotes1988chemotaxis,king2009chemotaxis,insall2010understanding,nichols2015chemotaxis}. A tremendous amount of research has built on this body of work, as reviewed in Refs.~\citenum{devreotes2003eukaryotic,jin2006moving,jin2008chemotaxis,swaney2010eukaryotic,insall2010understanding,roussos2011chemotaxis,levine2013physics}, to elucidate chemotaxis in diverse eukaryotes. For example, studies of human leukocytes have demonstrated that---similar to \textit{D. discoideum}---these cells respond to extracellular chemical stimuli by deforming and extending lamellipodia before migrating ~\citep{gerisch1981chemotactic,parent1998g}, a process driven by actin polymerization (Fig.~\ref{fig:figure6}\textbf{A}). Additionally, these cells are able to reorient and perform U-turns in response to an external chemical gradient. This behavior is distinct from chemotaxis of bacteria and other prokaryotic organisms, which instead perform biased random walks~\citep{Frymier:1995,Lauga:2006} as described in Section~\ref{bac-stimuli}.



Indeed, unlike prokaryotes, eukaryotic cells sense chemical gradients by measuring how the chemoattractant varies with respect to their size---a feat accomplished by triggering complex intracellular signals that drive cell polarization, deformation, and locomotion~\citep{jin2008chemotaxis,insall2010understanding,levine2013physics,thomas2018decoding}. In particular, chemotaxis in eukaryotic cells arises in three distinct steps---directional signaling, polarization, and motility---in which cells interact with the chemoattractant \com{or chemorepellent} and generate a response that determines the direction of motion \textit{via} intracellular signals. Several theoretical models have been developed to describe these signal transduction dynamics in \textit{D. discoideum} cells~\citep{meinhardt1999orientation,levchenko2002models,levine2006directional,levine2013physics}, resulting in local excitation-global inhibition (LEGI) equations that quantify the competition between the global chemical signal and the production of a local membrane-bound activator~\citep{parent1999cell}.

As described above, individual eukaryotic organisms employ chemotaxis for tracking food, finding the best path in complex environments such as soil, and escaping from predators~\citep{jin2006moving,jin2008chemotaxis,swaney2010eukaryotic,roussos2011chemotaxis,levine2013physics}. But chemotaxis is also used to guide collective cell migration, namely in \textit{in vivo} processes including metastasis~\citep{roussos2011chemotaxis}, tissue self-organization during morphogenesis~\citep{haas2006chemokine}, embryogenesis, immune surveillance, and wound healing. For example, chemotaxis enables immune cells such as T lymphocytes and neutrophils, which typically move randomly, to transition to a directed swarm behavior when inflammation occurs, in response to chemokine gradients~\citep{malet2015collective}.  Fig.~\ref{fig:figure6}\textbf{B} exemplifies this process: cells initially move randomly, but after a few hours, form multicellular clusters that sense a difference in chemokine concentrations across them and migrate toward a chemokine source. Similar behavior has been observed during tumor invasion of 3D hydrogels~\citep{liu2013minimization}. In particular, experiments~\citep{theveneau2010collective,malet2015collective} and theoretical modeling (e.g. agent-based, phase field, and cellular Potts models)~\citep{devreotes2003eukaryotic,camley2016emergent,najem2016phase} have shown that cells located at the \com{periphery of the colony} use the local information of the chemokine concentration to regulate their contact inhibition of locomotion, enabling the cluster directed migration. 

More recently, Ref.~\citenum{Tweedy2020} reported a combination of experiments and mathematical modeling showing that both \textit{D. discoideum} and metastatic cancer cells are able to solve complex mazes by using a self-generated chemotaxis mechanism that allows them to perform long-range navigation. This mechanism is in many ways similar to that by which bacterial populations perform chemotactic migration in response to a self-generated nutrient gradient, as described in Sec.~\ref{bac-stimuli}. In particular, Ref.~\citenum{Tweedy2020} showed that cells produce local chemical gradients in their vicinity, driving directed migration at essential decision-making points in the maze such as corners, junctions, shortcuts, and dead-end pores---suggesting a fundamental role of chemotaxis in regulating migration in complex environments during both healthy and pathological \textit{in vivo} processes. Also within the context of tumor invasion, Ref.~\citenum{roussos2011mena} explored the different types of cell motility that can arise during tumor cell invasion in a complex tissue environment in the presence of chemoattractants. In particular, the authors examined the actin and cell-migration regulatory protein called Mena and its isoforms Mena11a and Mena$^{\mbox{\tiny{INV}}}$, which are associated with breast cancer cell discohesion, invasion, and intravasation. They showed that Mena$^{\mbox{\tiny{INV}}}$ has higher sensitivity to epidermal growth factor than Mena11a, promoting coordinated and directed carcinoma cell streaming through the primary tumor, as shown in Fig.~\ref{fig:figure6}\textbf{C} (bottom). Moreover, they showed that such coordinated streaming migration is a long-lived phenomenon, as observed in Fig.~\ref{fig:figure6}\textbf{C} (top), persisting for longer than 24 hours.

As highlighted above, several theoretical works have incorporated chemotaxis into variants of agent-based and discrete models. Similarly accounting for chemotaxis in active-vertex models~\citep{barton2017active} and in continuum theories based on liquid crystal theory and active-gel physics~\citep{de1993physics,marchetti2013hydrodynamics,prost2015active,Alert2020} would also be interesting.\\



\begin{figure*}
  \begin{center}
        \includegraphics[width=.7\textwidth]{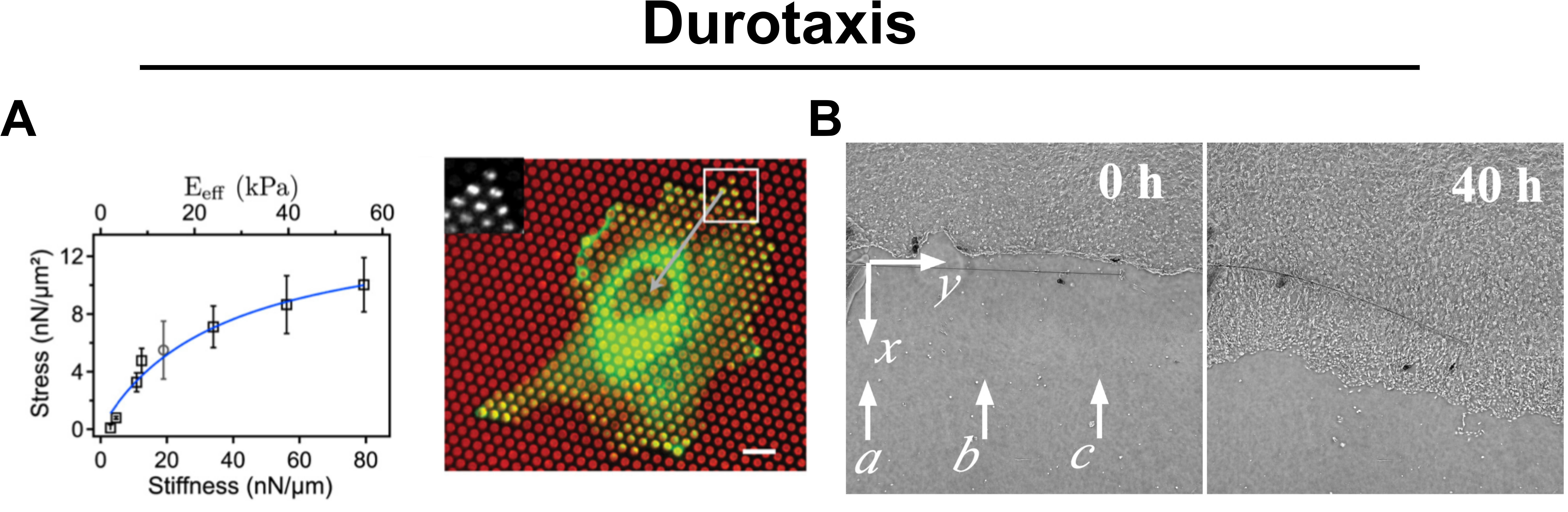}
    \caption{\begin{small}\textbf{Examples of durotaxis influencing both single and collective migration of eukaryotic cells}. \textbf{A}. Eukaryotic cell (REF52 fibroblast) spreading between on top of a surface of micropillars with varying rigidity, and the corresponding applied stress as a function of the substrate stiffness. Scale bar is $10~\mu$m. Reprinted with permission from Ref.~\citenum{trichet2012evidence} (National Academy of Sciences, 2012). \textbf{B}. Monolayer of human skin keratinocyte cells deforming a flexible carbon fibre. Scale bar is $1~$mm. Reprinted with permission from Ref.~\citenum{valencia2021interaction} (Elsevier, 2021).\end{small}\label{fig:figure7}}
  \end{center}
\end{figure*}

\noindent \textbf{Durotaxis.} As recently as 20 years ago, the authors of Refs.~\citenum{pelham1997cell,lo2000cell} discovered that eukaryotic cells are able to sense gradients in the stiffness of an underlying substratum. In particular, fibroblasts adapt their shape and move in the direction of increasing substrate stiffness, a phenomenon termed \textit{durotaxis}---derived from the Latin word \textit{durus}, meaning hard, and the Greek word \textit{taxis}, meaning arrangement. Since the original finding, research on durotaxis has remained active due to its potential implications in processes such as tissue regeneration, development, and tumor invasion. For example, studies have shown that diverse eukaryotic cells sense and respond to the rigidity of underlying substrata by changing shape and applying mechanical forces to probe the stiffness of their environment~\citep{discher2005tissue,saez2005mechanical,isenberg2009vascular,trichet2012evidence,vincent2013mesenchymal,roca2013mechanical,sunyer2020durotaxis}. The local mechano-sensitivity mechanism is usually explained in terms of force-induced conformational changes of constituent proteins at focal adhesion sites, which enhances the growth of local contacts by enabling the binding of new proteins~\citep{nicolas2004cell}. 

Other mechanisms by which cells can sense differences in substrate stiffness have also been suggested. Some authors, based on rearward actin cytoskeleton flow, have proposed a rigidity-dependent motor-clutch mechanism in which the strength of the molecular clutches linking F-actin to the substrate, and consequently the transmitted traction forces, depend on the substrate stiffness~\citep{chan2008traction,harland2011adhesion,bangasser2013determinants}. Other studies have shown that cells apply mechanical forces that keep the deformation of the substrate constant, thus maintaining larger traction forces in stiffer regions, thereby enabling migration in the direction of increasing substrate rigidity~\citep{saez2005mechanical,ghibaudo2008traction,ghassemi2012cells,trichet2012evidence} (Fig.~\ref{fig:figure7}\textbf{A}). This process has been proposed to be achieved either locally \textit{via} actin-myosin interactions at integrin-mediated focal adhesion sites~\citep{sheetz1998cell,fouchard2011acto,ghassemi2012cells,plotnikov2012force}, or at the global cell scale by cytoskeleton actin stress fibers pulling in focal adhesions~\citep{trichet2012evidence}. Intriguingly, there is an optimal stiffness range (3-10 nN/$\mu$m) in which durotaxis is most effective, with about 70\% of cells moving toward the stiffer region. Durotaxis can also be negative, with migration directed towards softer regions, for instance in retinal ganglion cell axons in \textit{Xenopus} developing brain~\citep{koser2016mechanosensing}. Indeed, the authors of Ref.~\citenum{isomursu2020negative} recently observed negative durotaxis in human glioblastoma (brain tumor) cells and in breast cancer cells (switched \textit{via} talin downregulation)---suggesting that solely integrin-mediated adhesion and motor-clutch dynamics are sufficient to explain both positive and negative durotaxis.

While initial research efforts focused on durotaxis of isolated cells, recent work has shown that groups of eukaryotic cells also sense substrate rigidity and migrate collectively in a preferred direction imposed by the stiffness gradient---a phenomenon termed \textit{tissue durotaxis}~\citep{sunyer2016collective}. In this work, the authors showed that 2D rectangular clusters of cells expand asymmetrically faster toward the stiffer region of the substrate. Their findings indicated that proliferation does not contribute to this asymmetric migration; instead, cell-cell adhesion is required for collective cell durotaxis. Thus, durotaxis is not only a consequence of local sensing of substrate rigidity, but also can involve a long-range mechanism mediated by adhesion between cells. This work also demonstrates that durotaxis is a robust guidance mechanism that can be used for tuning the directed collective motion of eukaryotic cells.




Durotaxis was firstly discovered in healthy cells, but it can also drive the 2D migration of cancerous cells, as shown in Refs.~\citenum{duchez2019durotaxis,sunyer2020durotaxis}. Nevertheless, to the best of our knowledge, collective durotaxis has not yet been demonstrated \textit{in vivo} in healthy or malignant cells~\citep{sunyer2020durotaxis}. Furthermore, it may be that cells use durotaxis as a migration cue to navigate through complex 3D tissue environments, such as when malignant cells leave a tumor and migrate through the surrounding extracellular matrix, a behavior associated with dissemination and metastasis. Indeed, there is a strong relationship between the stiffness of the environment and cancer cell spreading, which is mainly achieved \textit{via} cancer‐associated fibroblasts~\citep{bonnevie2018physiology}. Exploring these possibilities will be a useful direction for future work.

More generally, an important direction for future experiments is to explore other examples of cellular durotaxis in response to 2D and 3D deformable boundaries of complex shapes. However, fabricating substrata with well-defined shapes and stiffness gradients remains a challenge. Some experimental techniques that are often used involve preparing substrata by coalescing unpolymerized gel droplets of different concentrations and then polymerizing them; using microfluidic technologies to prepare substrata of different chemistries; and preparing substrata using photopolymerizable polymer mixtures~\citep{sunyer2020durotaxis}. In another approach, in recent work, the authors of Ref.~\citenum{valencia2021interaction} described a way to study the interaction of an epithelial monolayer of human skin keratinocyte cells with a deformable carbon fiber (Fig.~\ref{fig:figure7} \textbf{B}). This approach could pave the way to new experimental techniques able to measure the force exerted by a developing tissue, or to explain the interaction between groups of cells and external compliant bodies. Developing other ways to experimentally explore durotaxis will be a useful direction for future work.

Moreover, given that durotaxis is still an emerging area of research, there are only a few theoretical and computational studies that examine this mode of directed migration. As mentioned previously, some works have employed molecular clutch models to explain the underlying physics during durotaxis~\citep{chan2008traction,harland2011adhesion,bangasser2013determinants,sunyer2020durotaxis}. However, extending these clutch models to the case of collective cell durotaxis remains to be done. Within the context of continuum theories, the authors of Ref.~\citenum{alert2018role} recently analyzed the influence of durotaxis on the wetting/dewetting transition of a circular spreading tissue monolayer. The epithelial monolayer was modeled as a compressible polar fluid, taking into account cell-substrate friction, contact-active traction, and tissue contractility. The authors incorporated the substrate stiffness gradient by assuming that the cell-substrate friction and contact-active traction depend on the substrate elasticity. Their results suggest that tissue durotaxis can emerge from different wetting/dewetting states of the two moving edges of a tissue induced by the substrate stiffness gradient. For sufficiently large contractile stresses, the edge on the stiffer region spreads (wetting) and drags the other edge, causing it to retract (dewetting)---in agreement with the asymmetric migration towards the stiffer region observed in Ref.~\citenum{sunyer2016collective}. For small contractility, both edges wet the substrate, but the edge on the softer region slows down since moving towards decreasing stiffness decreases its active traction. More recently, by means of theory and numerical simulations within the same continuum framework as in Refs.~\citenum{blanch2017effective,alert2018role,Alert2020}, the work of Ref.~\citenum{trenado2021cellsubstrate} showed that solely cell-substrate friction is responsible for triggering complex secondary fingering instabilities at the edges of a 2D spreading monolayer, qualitatively similar to those observed in experiments. To test these predictions~\citep{alert2018role,Alert2019,trenado2021cellsubstrate}, new experiments must be conducted to systematically examine the ability of such active continuum models to describe experimental observations.


As a final note, we emphasize that eukaryotic cells respond to many other stimuli---for example through haptotaxis~\citep{carter1967haptotaxis,sengupta2021principles} (directed transport up a gradient in cellular adhesion sites or substrate-bound chemoattractants), phototaxis~\citep{foster1984rhodopsin,arrieta2019light} (directed transport in response to light), and topotaxis~\citep{park2018topotaxis,sengupta2021principles} (directed transport in response to spatial variations of topography). Indeed, multiple external stimuli often coexist in \textit{in vivo} processes together with a structured 3D environment. Hence, an important direction for future research is to elucidate the influence of multiple simultaneous stimuli on cellular migration. Although challenging, new experiments and both discrete and continuum theoretical models accounting for coupled external stimuli will shed light on different aspects of the \textit{in vivo} processes discussed throughout this chapter.

\section{Outlook}\label{sec:sec4}
This chapter has reviewed the transport of two prominent forms of biological active matter---bacteria and eukaryotic cells---in complex environments. Due to space constraints, we have not presented a fully comprehensive review of all the studies concerning it; instead, we have highlighted several sub-topics concerning which research has been particularly active. Within each section, we have described specific open questions and prospects for future research. Therefore, we conclude this chapter by highlighting general principles and areas for future inquiry. 

We have described ways in which---across both bacteria and eukaryotic cells---geometric constraints, mechanical properties, and external stimuli fundamentally alter active transport, modulating spreading and directed migration in often unexpected ways. Scientifically, these findings motivate the development of new models of transport. Furthermore, they challenge and expand current understanding of collective behavior and emergent phenomena in complex systems. Practically, these findings help predict and control a broad range of applications. For example, they underlie significant biomedical processes such as the progression of infection, drug delivery, tissue repair/wound healing, morphogenesis and development, and the progression of cancer. They also help inform the development of active systems that can rationally direct the transport of cargos, localize chemical transformations, or detect and respond to stimuli in environmental and biological settings \citep{Guix:2018,Alapan:2019,Erkoc:2019,Toley:2012,Duong:2019, Sedighi:2019,Soler:2013,Gao:2014,Adadevoh:2016,Ren:2019,Mano:2017,Gomez:2017,Vutukuri:2020,Nelson2010,Patra2013,Abdelmohsen2014,Wang2012a,Ebbens2016,soria2020predictive}.

While we have focused on the transport of bacteria and eukaryotic cells, extensive research has and continues to explore the ideas and phenomena described in this chapter for diverse other forms of active matter---ranging from other bacteria and eukaryotic cell types, as well as other microscopic organisms such as algae and protists, archaea, and sperm, subcellular active matter such as enzymes and driven biopolymer assemblies, and more macroscopic forms of active matter such as driven granular media, birds, fish, large mammals, and robots \citep{Bechinger2016b,Lang:2018,Deblais:2020phase,Deblais:2020rheology,Volpe:2017,soria2020predictive}. While these different forms of active matter vary across many orders of magnitude in size and speed, and have vastly different sources of activity, many of them share common features and therefore can often be described using more general models (e.g., Ref. \citenum{alert2021active}). An outstanding challenge is to develop a broad framework capable of quantitatively describing these non-equilibrium, yet universal, features---but also, to explore the limitations of such general descriptions and more deeply understand the rich phenomena that are specific to each system.

\section*{Acknowledgements}
A.M.-C. acknowledges support from the Human Frontier Science Program (LT000035/2021-C), from the Princeton Center for Theoretical Science, and from the FEDER/Ministerio de Ciencia, Innovaci\'{o}n y Universidades – Agencia Estatal de Investigaci\'{o}n through the project DPI2017-88201-C3-3-R, and the Red Nacional para el Desarrollo de la Microflu\'{i}dica, RED2018-102829-T. C.T. acknowledges financial support by the FEDER/Ministerio de Ciencia, Innovaci\'{o}n y Universidades – Agencia Estatal de Investigaci\'{o}n grant MTM2017-84446-C2-2-R, by the Madrid Government (Comunidad de Madrid-Spain) under the Multiannual Agreement with UC3M in the line of Excellence of University Professors (EPUC3M23), and in the context of the V PRICIT (Regional Programme of Research and Technological Innovation). S.S.D. acknowledges financial support by NSF grant CBET-1941716 and EF-2124863, the Project X Innovation Fund, the Eric and Wendy Schmidt Transformative Technology Fund at Princeton, the Princeton Catalysis Initiative, and the Princeton Center for Complex Materials, a Materials Research Science and Engineering Center supported by NSF grant DMR-2011750. Any opinions, findings, and conclusions or recommendations expressed in this material are those of the authors and do not necessarily reflect the views of the National Science Foundation.


\end{document}